\documentclass[fleqn,10pt]{wlscirep}
\usepackage[utf8]{inputenc}
\usepackage[T1]{fontenc}
 \usepackage[version=4]{mhchem}
\DeclareUnicodeCharacter{0394}{$\Delta$}
\newcommand{\blue}{\textcolor{black}}

\usepackage[normalem]{ulem}

\title{High Entropy Alloy under Shock Compression: Optical-Pump X-Ray-Probe } 

\author[1,+]{Hsin-Hui Huang}
\author[2,3,+]{Meguya Ryu}
\author[3]{Shuji Kamegaki}
\author[4,5]{Dominyka Stonyt\.{e}}
\author[6]{Tadas Malinauskas}
\author[7]{Yoshiaki Nishijima}
\author[8]{Rosalie Hocking}
\author[1]{Nguyen Hoai An Le}
\author[1]{Tomas Katkus}
\author[1]{Haoran Mu}
\author[1]{Soon Hock Ng}
\author[9]{Samuel Pinches}
\author[9]{Andrew S.M. Ang}
\author[9]{Christopher Berndt}
\author[10]{Martin Nicolaus}
\author[11]{Hans J\"{u}rgen Maier}
\author[10]{Kai M\"{o}hwald}
\author[12]{Vygantas Mizeikis}
\author[1]{Nadia Zatsepin}
\author[13]{Kohei Miyanishi}
\author[13,14]{Toshinori Yabuuchi}
\author[15]{Hirotaka Nakamura}
\author[15]{Alexis Amouretti}
\author[15,16]{Norimasa Ozaki}
\author[17]{Tommaso Vinci}
\author[18,19]{Arturas Vailionis}
\author[20]{Eugene G. Gamaly}
\author[1]{Damien G. Hicks}
\author[2,3,21,22]{Junko Morikawa}
\author[1,5,21]{Saulius Juodkazis}

\affil[1]{Optical Sciences Centre and ARC Training Centre in Surface Engineering for Advanced Materials (SEAM), School of Science, Swinburne University of Technology, Hawthorn, Victoria 3122, Australia}
\affil[2]{
School of Materials and Chemical Technology, Institute of Science Tokyo, 2-12-1, Ookayama, Meguro-ku, Tokyo 152-8550, Japan}
\affil[3]{CREST-JST and Institute of Science Tokyo, Meguro-ku, Tokyo 152-8550, Japan}
\affil[4]{Direct Machining Control, UAB, Mokslininku st. 6B, 08412 Vilnius, Lithuania}
\affil[5]{Laser Research Center, Physics Faculty, Vilnius University, Saul\.{e}tekio Ave. 10, 10223 Vilnius, Lithuania}
\affil[6]{Institute of Photonics and Nanotechnology, Vilnius University, Lithuania}
\affil[7]{Department of Electrical and Computer Engineering, Graduate School of Engineering, Yokohama National University, 79-5 Tokiwadai, Hodogaya-ku, Yokohama, 240-8501, Japan}
\affil[8]{Chemistry, Swinburne University of Technology, John St, Hawthorn, 3122 Vic, Australia}
\affil[9]{Australian Research Council (ARC) Industrial Transformation Training Centre in Surface Engineering for Advanced Materials (SEAM), Swinburne University of Technology, Hawthorn, VIC, 3122, Australia}
\affil[10]{Joining and Surface Technology Division, Institute of Materials Science, Leibniz Universit\"{a}t Hannover, Witten, Germany}
\affil[11]{Institute of Materials Science, Leibniz Universit\"{a}t Hannover, Hannover, Germany}
\affil[12]{Shizuoka University, Research Institute of Electronics, Hamamatsu, Japan}
\affil[13]{RIKEN SPring-8 Center, Sayo, Hyogo 679-5148, Japan}
\affil[14]{Radiation Research Institute, Sayo, Hyogo 679-5198, Japan} 
\affil[15]{Graduate School of Engineering, University of Osaka, Suita, Osaka 565-0871, Japan}
\affil[16]{Institute of Laser Engineering, University of Osaka, Suita, Osaka 565-0871, Japan}
\affil[17]{\'{E}cole Polytechnique, Palaiseau, Laboratoire pour l’utilisation des lasers intenses (LULI), CNRS UMR 7605, Route de Saclay, 91128 Palaiseau Cedex, France }
\affil[18]{Stanford Nano Shared Facilities, Stanford University, Stanford, CA 94305, USA}
\affil[19]{Department of Physics, Kaunas University of Technology, LT-51368 Kaunas, Lithuania}
\affil[20]{Laser Physics Centre, Department of Quantum Science and Technology, Research School of Physics, The Australian National University, Canberra, ACT 2601, Australia}
\affil[21]{World Research Hub (WRH), School of Materials and Chemical Technology, Institute of Science Tokyo, 2-12-1, Ookayama, Meguro-ku, Tokyo 152-8550, Japan}
\affil[22]{Research Center for Autonomous Systems Materialogy (ASMat), Institute of Innovative Research, Institute of Science Tokyo, Yokohama 226-8501, Japan}
\affil[+]{these authors contributed equally to this work}
\begin{abstract}
High entropy alloys (HEAs) are multi-principal-element alloys designed for tailorable mechanical performance and have been attracting significant engineering interest, yet their fundamental behaviour under extreme dynamic conditions, such as shock loading, remains unexplored. Here, we report laser-shock experiments on two different types of $\sim 1$-$\mu$m-thick HEA microfilms, CuPdAgPtAu and CrFeCoNiCuMo, on $\sim 25$-$\mu$m-thick black-Kapton ablator driven by a high intensity laser pulse (532~nm, 5~ns, 16~J, $\sim 0.5$-mm diameter focal spot) and probed by an X-ray free electron laser (XFEL) pulse (12~keV, 7~fs). Time-resolved X-ray diffraction (XRD) shows the formation of a transient phase with a lattice compression up to $\sim 5.1\%$ of the CuPdAgPtAu HEA along the (111) plane; 
this transient compressed phase existed for $\sim 0.3$~ns. 
The impedance matching Hugoniot analysis estimated a shock pressure of $55\pm 6$~GPa in the HEA film, while Au- and Fe-based equations of state (EoS) modelling predict 80~GPa (0.8~MBar) at the free HEA surface. The free HEA surface reached maximum velocities of $\approx 5$~km/s as recorded from \emph{in situ} monitoring with the velocity interferometry system for any reflector (VISAR) imaging. These initial HEA results show the suitability of HEA sample preparation and XFEL-based XRD characterisation under extreme shock loading, and are promising for experimental determination of the EoS of this emerging class of materials (beamtime proposal No.: 2024A8503 for a 6-hour preliminary experiment).
\end{abstract}
\begin{document}

\flushbottom
\maketitle
%
%
\tableofcontents
\begin{figure}[ht!]
\centering\includegraphics[width=\textwidth]{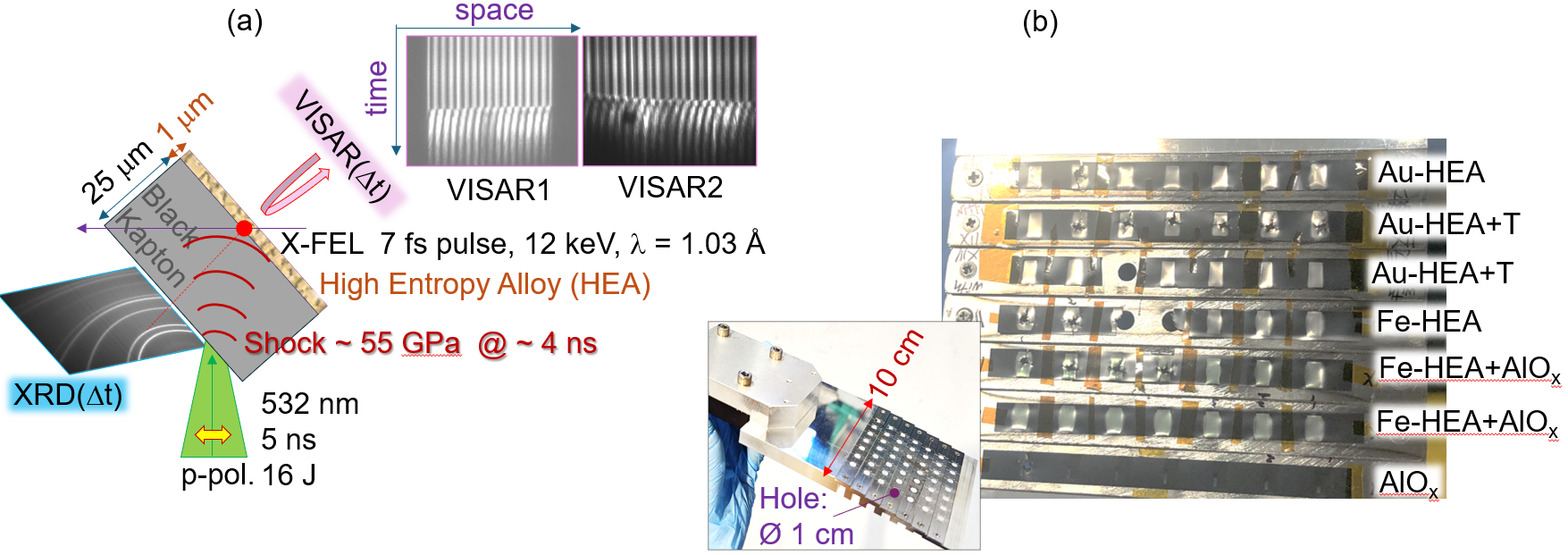}
\caption{Setup used in the 2024A8503 SACLA beamtime; all capabilities are described in ref.~\citenum{BL3}. (a) Geometry of experiment with an XRD flat-panel detector (FPD) at the bottom (transmission geometry), and two VISARs (velocity interferometer system for any reflector) at 532~nm wavelength. (b) Mounted samples after the experiment. The inset shows a $7\times 7$ array of 1-cm-diameter holes for samples. HEA samples of CuPdAgPtAu (Au-HEA) and CrFeCoNiCuMo (Fe-HEA) were deposited as $1~\mu$m-thick film on a $25~\mu$m-thick black Kapton ablator. Annealed samples are denoted ``+T'' and alumina-coated samples are shown with \ce{AlO$_x$} ($\sim 0.8~\mu$m film for better VISAR visualisation and reduced shock reflection). One hole was left without a sample for XFEL beam alignment. Samples of alumina film on black Kapton had gold patches sputtered at a corner for alignment. 
 }
\label{f-setup}
\end{figure}
\begin{figure}[tb]
\centering\includegraphics[width=0.75\textwidth]{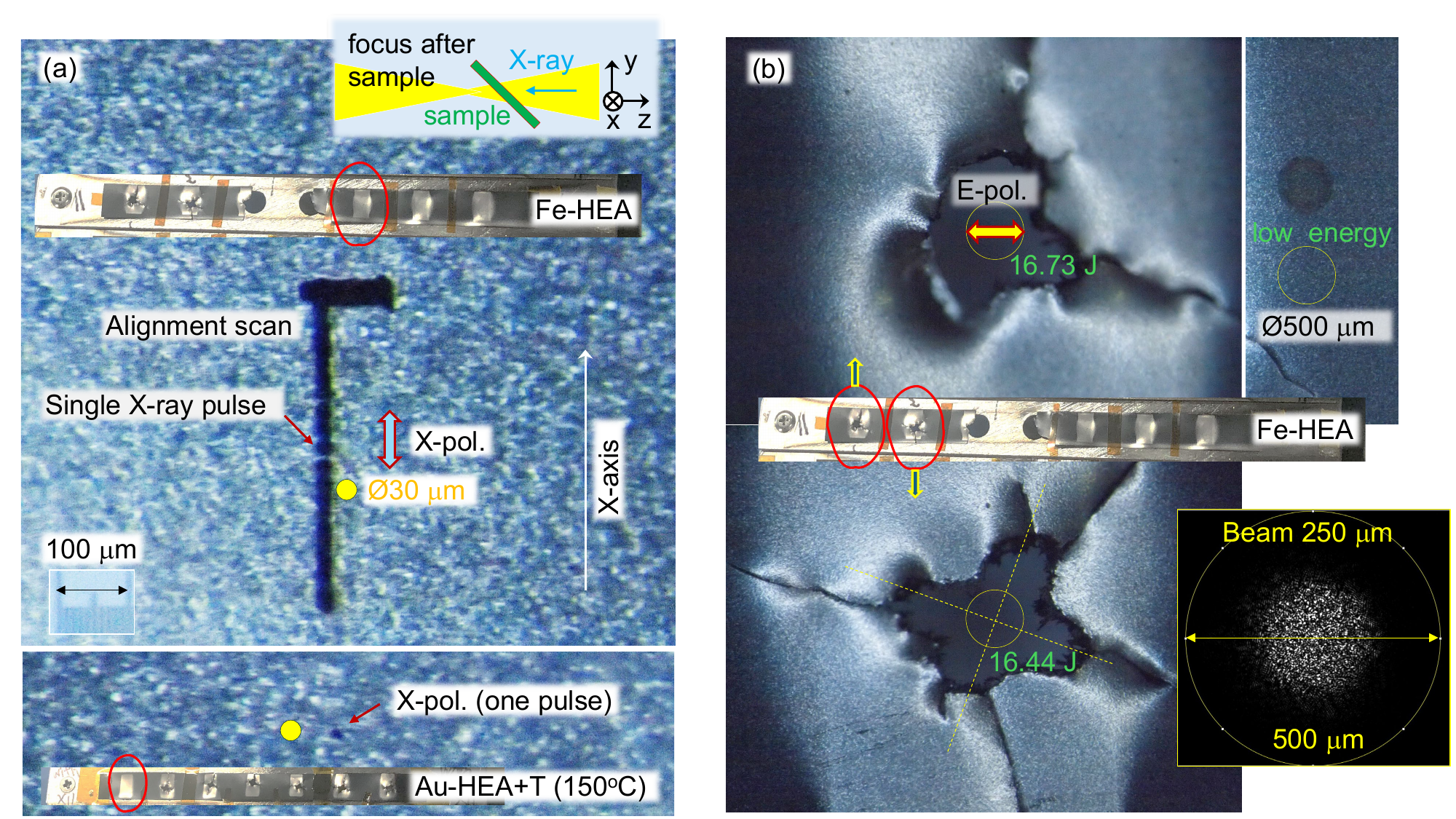}
\caption{Optical images of (a) X-ray induced ablation on the Fe-HEA film on black Kapton and (b) damage from a $\sim 16$~J, 532~nm, 5~ns optical pulse through the Fe-HEA film on black Kapton (top-view from the HEA side). Insets in (b) show low-energy optical pulse damage on the top surface of the HEA film and an optical image of an approximately 250~$\mu$m diameter (FWHM) top-hat beam (half the $\sim$500~$\mu$m full beam diameter, courtesy of the BL3 beamline team). A dashed cross at the bottom image in (b) shows the orientation of the DOE, which was used to generate the top-hat pump beam. } 
\label{f-shot}
\end{figure}

\begin{figure}[h!]
\centering\includegraphics[width=.85\textwidth]{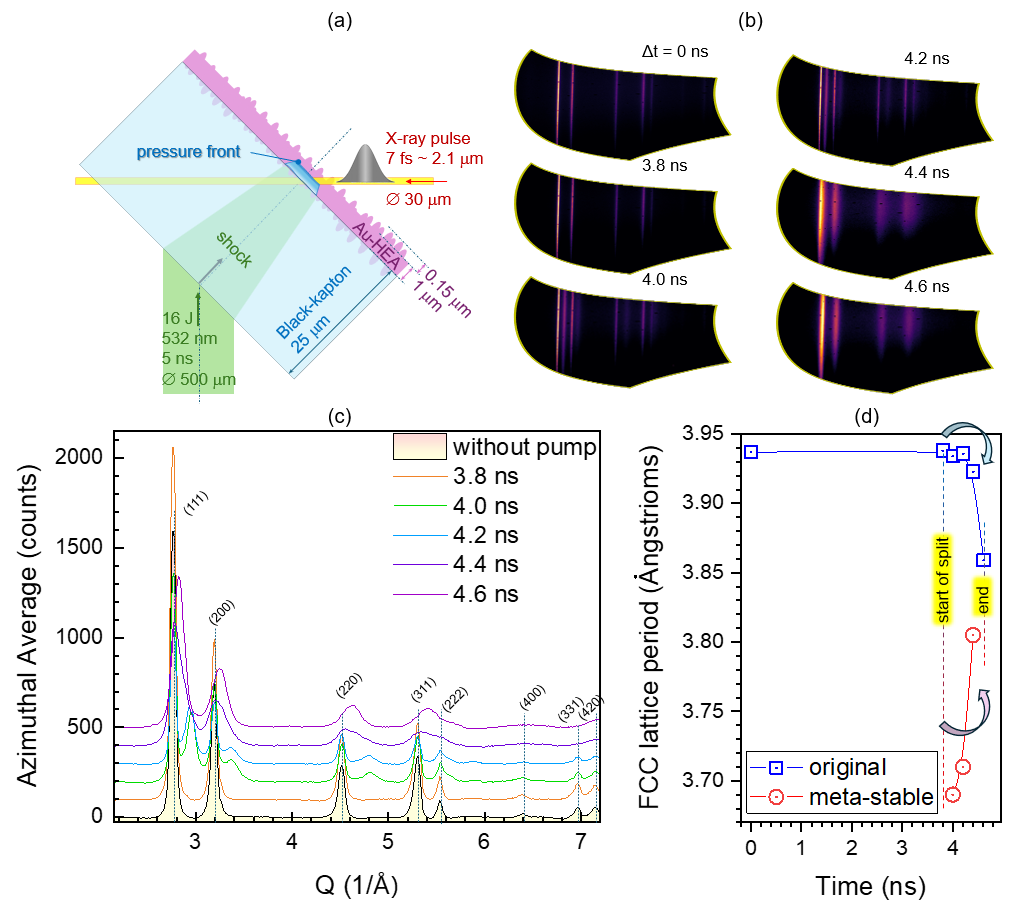}
\caption{The structural evolution upon the application of shock wave probed by high-speed XFEL-based XRD structural analysis of CuPdAgPtAu HEA (Au-HEA) film: (a) geometry of interaction region, (b) azimuthally integrated $2\theta$-angular XRD images and (c) 
integrated profiles vs. $Q$~[$\text{\AA}^{-1}$]; see Fig.~\ref{f-AuSlice} for the side-view microtome cross-section. Simulations using Crystal Maker/Diffract for 9-nm FCC structures are shown \blue{is this 9 nm the black lines in file "f-auxrd1" for different lattice sizes?}. The surface of Au-HEA coating exhibits a random pattern of protrusions up to 150~nm in height (as observed by SEM and AFM). (d) The Pawley analysis of the lattice period evolution showing a split into two phases, which converge into a single and more disordered phase(see text for discussion). }
\label{AuHEA}
\end{figure}

\begin{figure}[tb]
\centering\includegraphics[width=0.75\textwidth]{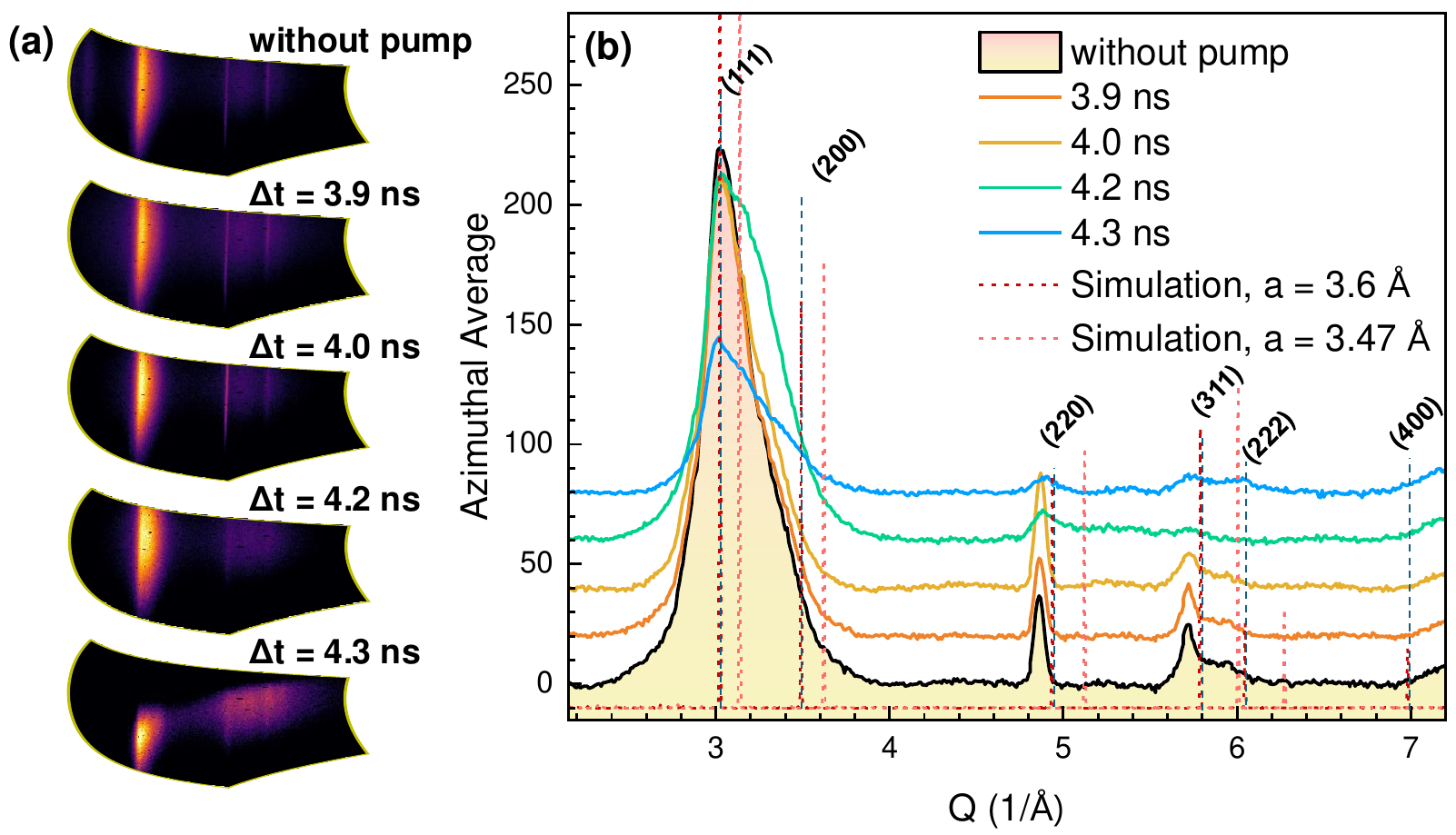}
\caption{The structural evolution upon the application of 
\blue{pressure} wave probed by high-speed XFEL-based XRD structural analysis of a $\sim 1~\mu$m CrFeCoNiCuMo HEA film: (a) azimuthally integrated $2\theta$-angular XRD images and (b) integrated XRD $Intensity~vs.~Q$~[$\text{\AA}^{-1}$] plotted with vertical offset. See Supplement for the bandwidth changes during 
\blue{pressure} loading. 
}
\label{FeHEA}
\end{figure}

\section{Introduction}

The study of matter under extreme conditions is a frontier of modern science, advancing along with the development of extraordinary infrastructure such as high peak brilliance X-ray free electron lasers (XFELs)~\cite{Glenzer_2016,Fukuzawa_2020,Rastogi}. Focusing of a 7~fs 9.1~keV pulse into a 7~nm spot achieves intensities of $1.45\times 10^{22}$~W/cm$^2$ which take all electrons from a Chromium atom producing Cr$^{24+}$ state - bare nucleus~\cite{NP24}. This formidable result is achieved by a combination of convex and concave reflective Kirkpatrick-Baez (KB) mirrors at the grazing-incident to minimise aberrations which is achieved over $\sim 147$~m of X-ray optics~\cite{Cr24}. Such high intensities can be also achieved at visible or near-IR spectral range by ultra-short pulsed lasers using the optical parametric chirped pulse amplification (OPCPA)~\cite{Dubietis}. Terawatt ($10^{12}$~W) class lasers have tens-of-mJ pulse energy at $\sim 10$~fs duration and even higher Petawatt pulse powers are achievable by thin thermoplastic film post-compression technique~\cite{Mou22}. 
Lasers of Joule-level energies with pulse duration of tens-of-femtoseconds have higher repetition rates and deliver intensities up to $10^{20}$~W/cm$^2$ per pulse, when focused on a cylindrical wire delivering a converging 9-fold compression at the wire's axis reaching 800~Mbar pressure as were probed with femtosecond XFEL pulses~\cite{Laso}. These conditions are relevant to astrophysical phenomena and warm dense matter (WDM) research~\cite{wdm}.\\

It soon will become possible to produce practical 1~kHz and $10^{10}$~n/s neutron laser-based sources using high-intensity ultra-short pulses as demonstrated in the proof of the concept experiments at a scaled-down version with 1~Hz repetition rate 12~fs, 21~mJ pulses focused to $\sim 3~\mu$m diameter and reaching $4\times 10^{18}$~W/cm$^2$ intensity~\cite{Osvay}. This opens widely anticipated possibilities of nuclear transmutation of chemical elements and isotopes with unprecedented flexibility and versatility. For example, the currently used neutron activation of Si in nuclear reactors, can be achieved with simpler setups, i.e., n-type doping of Si by Phosphorus in a pure Si (a perfect lattice) is achieved via $^{30}_{14}$Si + n $\Rightarrow$ $^{31}_{14}$Si + $\gamma$ $\Rightarrow$ $^{31}_{15}$P + $\beta^-$(electron). Once relativistic intensities are reached above $I_l\sim 4\times 10^{18}$~W/cm$^2$, neutron production yield scales as $I_l^4$ as demonstrated using 1.5~ps 900~J pulses~\cite{Yogo}. Ultra-fast laser-driven MeV electron and neutron sources is a fast emerging field~\cite{Gunter}. 
The long-time aim to harness laser-driven fusion is also getting closer as the Lawson criterion for ignition was recently exceeded in a laser-driven inertial fusion experiment~\cite{Sun}. 

The creation of a high pressure with ultra-fast thermal quenching using fs-laser pulses opens a new synthesis pathway to new materials and phases, including metastable which can be retrieved to the ambient conditions~\cite{06prl166101}. Open-packed incommensurate host-guest (same or impurity atoms) structures are formed at high pressure, e.g., for Ba at 10~GPa~\cite{Nelmes}. A twofold role of guest atoms is manifested: 1) allowing for the formation of diverse structures having multiple enthalpy minima at a certain pressure, 2) enhancing the high-pressure phase stability, e.g., by implantation of guest atoms N, O. Aluminium forms the host-guest incommensurate structure at 3.2~TPa~\cite{Pickard,McM}. High-pressure Si phases were observed under ultra-relativistic $7.5\times 10^{19}$~W/cm$^2$ intensity using $\sim 30$~fs pulses for irradiation of Si and generation energetic MeV electrons which contributed to the strong back-side ablation of $\sim 0.5$~mm wafer~\cite{24prr023101}. 

Interaction of ultra-short laser pulses of high intensity with high entropy alloys (HEA) is the active focus area due to the promise of superior mechanical properties of HEAs at extreme conditions~\cite{ZHANG20221,alagarsamy_mechanical_2016,guo_microstructure_2022}. 

Analysis of data from 2024A8503 SACLA beamtime is presented here. We used high entropy alloys (HEAs) on black Kapton ablator shocked with optical pump beam 532~nm wavelength, $t_p = 5$~ns pulse duration focused onto $2r = 470~\mu$m (at FWHM) diameter top-hat spot; the beam was formed using a diffractive optical element (DOE). Pulse energy was $E_p = 16$~J, which corresponded to the fluence $F_p = E_p/(\pi r^2)\approx 9.2$~kJ/cm$^2$ and the average intensity $I_p=F_p/t_p \approx 1.84$~TW/cm$^2$ on the black Kapton surface. Typical shock travel time through the black Kapton (polyimide) to reach the HEA film was $\sim 4$~ns. A time-delayed X-ray pulse of XFEL 12~keV/7~fs ($2\times 10^{11}$~photons/pulse; 400~$\mu$J) was probing the shock-affected region on a 1-$\mu$m-thick HEA film with an elliptical focus of cross-sections $30\times 20~\mu$m$^2$. Single laser pulse was used for modification, which was probed at a few different delays by XFEL pulse. X-ray diffraction (XRD) was detected in transmission geometry (Fig.~\ref{f-setup}(a)). Two VISAR systems monitored the top surface of the HEA film and provided a time-resolved velocity measurement based on laser interferometry. Two VISAR systems were allowed to properly calibrate the temporal-spatial response of the surface to the applied shock. Technical capabilities of the used SACLA beamline can be found elsewhere~\cite{BL3}.

\section{Experimental: materials, methods, conditions}
\subsection{High Entropy Alloy (HEA)}

High entropy alloys (HEAs) are new alloy materials that were theorized~\cite{Tsai03} 
then realized~\cite{yeh_nanostructured_2004} around the early 2000s. Unlike traditional alloys, HEAs are composed of five or more primary elements, each with atomic concentrations ranging from 5\% to 35\%~\cite{yeh_nanostructured_2004,yeh_high-entropy_2007}. The elements used are usually 3d transition metals 
similar atomic size that leads to the tendency to form a solid solution when the configuration entropy is met.
\blue{
} HEAs are versatile due to their potential combinations. This opens the door for a wide range of applications such as hydrogen storage~\cite{kao_hydrogen_2010,kunce_structure_2013}, biomaterials~\cite{castro_overview_2021}, functional coatings~\cite{tsai_high-entropy_2014,li_review_2019,meghwal_thermal_2020}, nuclear applications~\cite{shi_current_2021,pickering_high-entropy_2021}, and more. With the potential of applying HEAs under harsh environments, it is important to understand the metamorphism of such materials under extreme conditions. Two different HEAs are selected for this experiment, CuPdAgPtAu (Au-HEA) and CrFeCoNiCuMo (Fe-HEA), due to their unit cell structures, range of lattice parameter differences between the constituent elements, and atomic size differences between the constituent elements. 
The measured lattice parameters of Au-HEA's constituent metals, determined with Cu-$K_\alpha$ 8.04~keV/1.54056\AA~ line~\cite{auhea}, are 4.08\AA~(Au, fcc), 4.09\AA~(Ag, fcc), 3.62\AA~(Cu, fcc), 3.89\AA~(Pd, fcc), and 3.90\AA~(Pt, fcc); and 2.51\AA~(Co, hexagonal structure(hex)), 2.88\AA~(Cr, body-centre cubic(bcc)), 2.87\AA~(Fe, bcc), 3.52\AA~(Ni, fcc), and 3.15\AA~(Mo, bcc) for Fe-HEA~\cite{Hermann_2011}. Both alloys have face-centered cubic (FCC) structure, with a theoretically predicted lattice parameter of $a=3.92$\AA~\cite{Miracle_2017} for Au-HEA and $a = 3.56$\AA~\cite{Miracle_2017} for Fe-HEA (3.59\AA~for CrFeCoNiCuMo), the theoretical parameters are calculated using the rule-of-mixtures approach 
$r_{alloy}\,=\, \mathop {\sum}\limits_{{\rm{i}}} {r}_{{\rm{i}}}{x}_{{\rm{i}}}$, 
where $\rm{i}$ is each constituent element, ${r}_{{\rm{i}}}$ is the radius of each constituent metals and the lattice parameter is calculated base on the FCC crystal structure, $a_{fcc} = 2 \sqrt{2} r$. The atomic radius for each constituent metals are 144.2 pm~(Au) 144.47 pm~(Ag) 127.8 pm~(Cu), 138.7 pm~(Pt), and 137.54 pm~(Pd), the atomic sizes distributed evenly with a maximum difference of 16.67~pm for Au-HEA. In the case of Fe-HEA, the atomic size are not as evenly distributed, 125.10 pm~(Co), 124.91 pm~(Cr), 124.12 pm~(Fe), 124.59 pm~(Ni), and 136.26 pm~(Mo), with only Mo being noticeably larger and a maximum difference of 12.14~pm for Fe-HEA~\cite{Miracle_2017}.
Base on the rule-of-mixtures approach, theoretical melting temperature can also be calculated with 
$T\,=\, \mathop {\sum}\limits_{{\rm{i}}}{T}_{{\rm{i}}}{x}_{{\rm{i}}}$,
where $\rm{i}$ is each constituent element, $x_i$ is the molar composition of each constituent metals, and $T_i$ is the melting point for the corresponding metals. The estimated melting point for Au-HEA is 1559.8K and 1985.55K for Fe-HEA.

The atomic weight for Au-HEA is estimated to be 133.98~g/mol~\cite{DeLaeter_2003} and the mass density is calculated to be 14.79~g/cm$^3$ base on $\rho = \frac{Z \times M}{N_A \times V_c}$ where $Z$ is the number of atoms per unit cell, $M$ is the molar mass in g/mol, $N_A$ is Avogadro's number, and $V_c$ is the volume of the unit cell which equals to $a^3$ for the FCC structure. Fe-HEA's atomic weight is estimated to be 60.77~g/mol~\cite{DeLaeter_2003} (CrFeCoNiCuMo has the atomic weight of 64.14~g/mol) and the mass density is calculated to be 8.97~g/cm. Thermal diffusivity of Fe-HEA was directly measured by Temperature Wave Analysis (TWA)~\cite{Morikawa} $\chi = 4.36\times 10^{-7}$~m$^2$/s. For the typical shock wave velocity of $v_{sh}\sim 6$~km/s in experiments for the $d = 1~\mu$m thickness of Fe-HEA, the heat diffusion length is negligible $\sqrt{\chi d/v_{sh}} = 8.5$~nm. 

\subsection{Sample Deposition}\label{sample_depo}

The Fe-HEA coupon was produced at FORTIS (Joining and Surface Technology) in Witten, Germany, 
via cold spray deposition onto copper sputter targets at 38~bar using a CGT Kinetics 4000/34. 

Both types of HEAs described above are used separately for the preparation of five different samples with physical vapour deposition method (PVD) or electron beam deposition method (e-beam, AXXIS, Kurt J. Lesker company, Jefferson Hills, PA, USA) onto a 25$~\mu$m thin Kapton B polyimide \ce{[-C6H4O-C6H4N(OC)2(C6H2)(OC)2N-]$_n$} film (DuPont, Wilmington, DE, USA). They were 
CuAgAu and PtPd high entropy alloy (HEA) targets co-sputtered by PVD to form $\sim 1~\mu$m of Au-HEA; the targets were purchased from Tanaka Precious Metal Co., Ltd., Japan. 
Similarly, a $\sim 1~\mu$m of CrFeCoNiCuMo (Fe-HEA) was sputtered by PVD. Furthermore, another set of $\sim 1~\, $m of CrFeCoNiCuMo HEA was sputtered with PVD with an additional $\sim 1~\mu$m layer of Aluminium oxide (AlO$_x$) deposited with e-beam for some samples to reduce shock as well as optical reflectance. 
The deposition rate of CuPdAgPtAu HEA (Au-HEA) is estimated to be 0.1865~$\mu$m/min/W ($7.46~\mu$m/min for the used 40~W) from target-1 CuAgAu and 0.1284~$\mu$m/min/W (5.13~$\mu$m/min at 40~W) from target-2 PdPt; Ar gas flow was at 20 sccm. The rotation speed of the sample during sputtering was 660~rpm. The deposition rate of CrFeCoNiCuMo HEA (Fe-HEA) and AlO$_x$ were $\sim$6~nm/min. 
All samples underwent extreme shock compression by 16~J laser pulses. The high-speed dynamics of their crystalline structure transformation were monitored with X-ray Free Electron Laser (XFEL) at various delay times from the shock wave generation. 

\subsection{SPring-8 Angstrom Compact free electron LAser (SACLA) beamline BL3}

The experiment was performed at the hard X-ray beamline in SACLA~\cite{sacla} Japan (BL3)~\cite{BL3} which is used for analysis of phase transitions under shock compression: cavitation melting~\cite{Liquid}, olivine-ringwoodite transformation at $\sim 23$~GPa (corresponding to 660~km inside Earth)~\cite{olivine}, stishovite at 300~GPa~\cite{acoustic}. The samples, $\sim 1~\mu$m thick Au-HEA or Fe-HEA on a 25~$\mu$m thick black Kapton polyimide film prepared as Sec.~\ref{sample_depo} has described, were shocked with a pump laser (532~nm, 5~ns, $\sim$16~J) focused to a $\sim 0.47$~mm diameter spot on the back side of the black Kapton film. The probing X-ray free electron laser (XFEL) beam, set at $\Delta t$ after the pump laser, was at 12~keV with a mean pulse energy of 370~$\mu$J/pulse (number of photons $N_X = 1.924\times 10^{11}$), and pulse duration of 7~fs (2.1~$\mu$m in space). The number density of photons in the focal region assumed as a cylinder 2.1~$\mu$m long and with a base of 15~$\mu$m in radius is $n_X = 1.3\times 10^{20}$~photons/cm$^{3}$ (photons per area $2.7\times 10^{16}$~photons/cm$^2$). As an estimate, X-ray fluence considering $\sim 15~\mu$m radius of the focal spot was $F_X = 52.4$~J/cm$^2$ and intensity $I_X = 7.5$~PW/cm$^2$. The X-ray beam sizes at the sample position were measured by the knife (wire) edge scanning. The horizontal (x-axis) diameter $d_x = 27.9\pm 1.9~\mu$m ($\pm\sigma$ standard deviation) at FWHM and vertical $d_y = 18.9\pm 1.2~\mu$m without considering the projection angle of $\theta_S = 45^\circ$. Considering sample inclination, the vertical size on the surface (a projection) becomes larger $d_y/\cos\theta_s\approx 27~\mu$m. X-ray polarization was in the horizontal p-pol. 

\subsection{X-ray Diffraction - XRD}

A flat panel detector (FPD) was used for direct digital radiography of 2D XRD. Shock compression causes a peak shift/splitting by an internal stress or planar faults, especially stacking faults or twinning. The peak broadening is related to the size of crystallites and micro-stresses. Also, the stress gradients and/or chemical heterogeneities can cause peak broadening. Similarly, the peak asymmetries can be caused by long-range internal stresses, planar faults or chemical heterogeneity~\cite{DEVINCENTIS,cryst8}. 
Moreover, an anisotropic peak broadening can result from an anisotropic crystallite shape or anisotropic strain~\cite{ungar_meaning_2003}.

XRD intensity in Fig.~\ref{FeHEA}(b) was obtained by azimuthal integration of FPD detector using standard procedure by the program heXRD~\cite{Boyce_2013,avery_2024}. In order to compare XRD data measured with different sources, a Q-spacing presentation was selected to display XRD data, see Fig.~\ref{FeHEA}(b). The scattering vector $Q = \frac{2\pi}{d} \equiv \frac{4\pi}{\lambda}\sin\theta$ is calculated using X-ray wavelength $\lambda$ and diffraction the angle $2\theta$; the Bragg spacing $d = \frac{n\lambda}{2\sin\theta}$, where $n$ is an integer number. \ce{CeO2} was used for calibration of diffraction angle. The Pawley method, which does not require prior knowledge of the structure, was used for XRD data analysis~\cite{Pawley}.


\begin{figure}[tb]
\centering\includegraphics[width=0.7\textwidth]{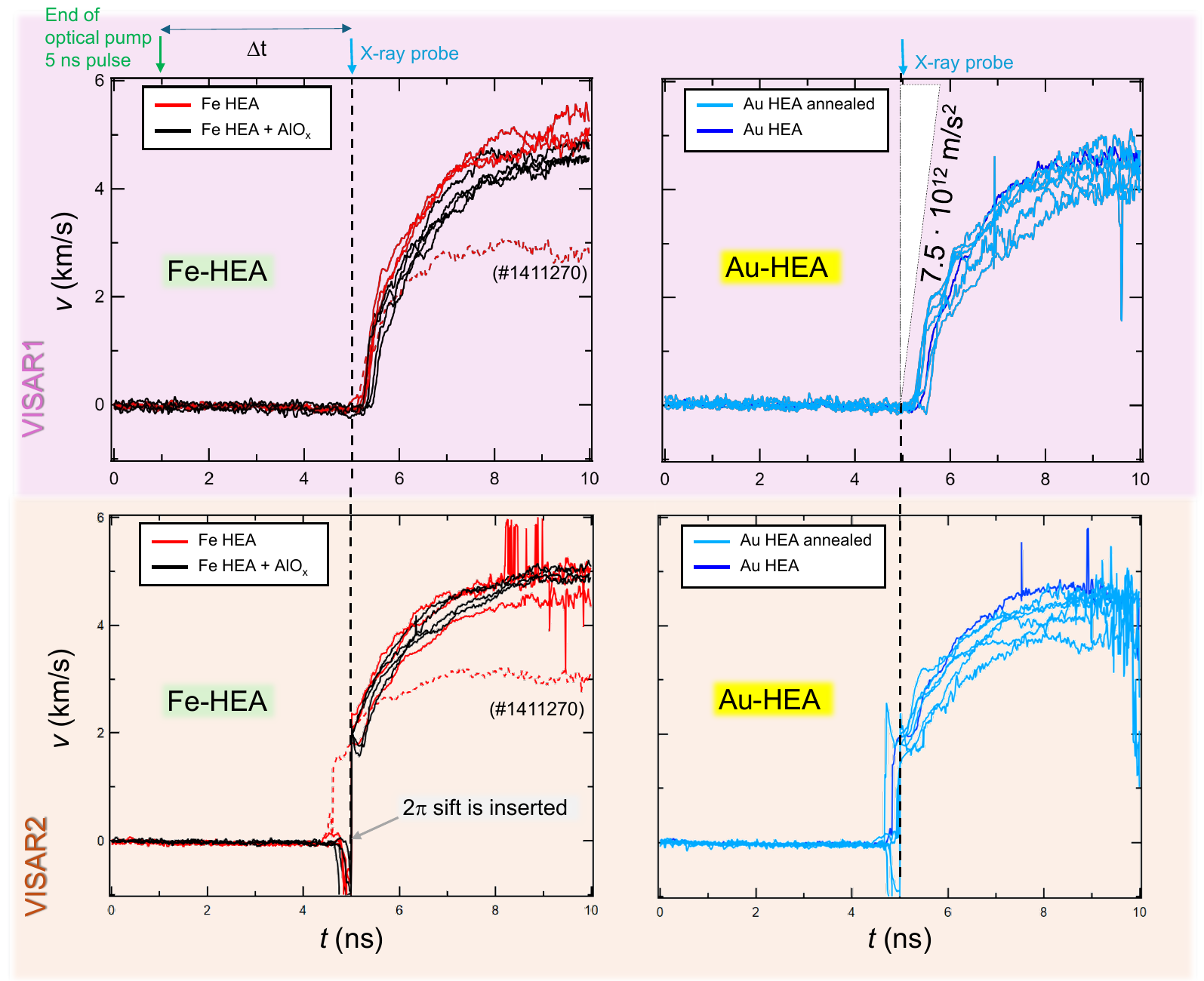}
\caption{Velocity vs. time from VISAR1 and VISAR2 data sets for Fe-HEA and Au-HEA samples. Each line corresponds to a single pulse of optical-pump pulse $\sim 16$~J; those lines are dots in Fig.~\ref{f-2vis}. At the negative acceleration point in VISAR2, a $2\pi$ phase was added. This resulted in a good correspondence between 
\blue{pressure wave} velocities reached at 9-9.5~ns $\sim 5$~km/s for both detectors. The shot number $\#1411270$ corresponded to approximately half of the interferogram (and partial circles of the XRD pattern), most probably due to the edge of film/sample. The $\sim$5~ns optical pump finishes at approximately 1~ns VISAR timer position; see Fig.~\ref{f-cond}(a) for an oscillogram of the optical pump pulse. The initial acceleration reached $\sim 7.5\times 10^{12}$~m/s$^2$ for Au-HEA. Laser pulse jitter is $\sim 160$~ps; uncertainty in determination of the instant velocity was $\pm 0.25$~km/s.}
\label{f-visar}
\end{figure}
\begin{figure}[tb]
\centering\includegraphics[width=0.4\textwidth]{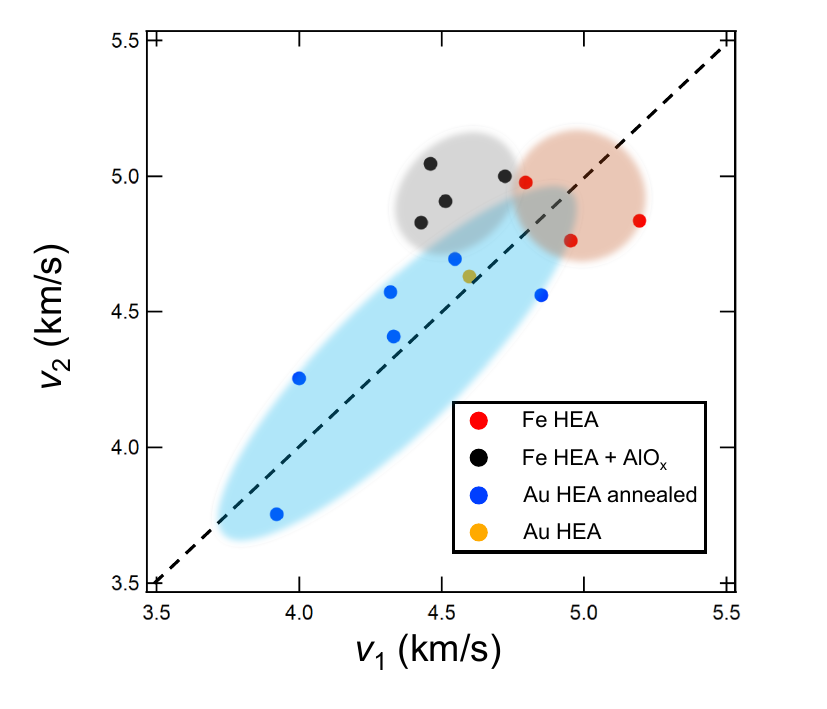}
\caption{Correlation of the 
\blue{pressure wave} velocities $v_1$ vs. $v_2$ at 9-9.5~ns time window measured with VISAR1 vs. VISAR2, respectively, from the same event (based on Fig.~\ref{f-visar}); uncertainty in determination of the instant velocity was $\pm 0.25$~km/s.}
\label{f-2vis}
\end{figure}

\subsection{Velocity Interferometry System for Any Reflector - VISAR}

Surface reflection from the HEA side was analysed using the imaging velocity interferometry system for any reflector (VISAR)~\cite{visar} (Fig.~\ref{f-setup}(a)); wavelength of illumination was 532~nm. One dimensional Fourier transform was performed along the space direction (horizontal direction of the image from streak camera in Fig.~\ref{f-setup}(a)) at each time step, and the spectrum in a limited frequency window is used for the inverse Fourier transform to extract the phase information of the fringes. The extracted phase is unwrapped by detecting the $2\pi$ phase shift in the phase profile.

Two interferometer systems having different sensitivity were monitoring the surface motion of HEA in parallel to extend the dynamic range of the velocity detection. The sensitivities of the VISAR1 streak camera was 5.419~km/s and VISAR2 2.894~km/s depending on the etalon thickness of the interferometer. In a case of Doppler shift of free surface in a vacuum condition, the velocity per fringe (VPF) can be written as: ${\rm VPF} = \frac{\lambda}{2\tau(1+\delta)}$, where $\lambda$ is the wavelength of the light for VISAR system, $\tau$ is the time delay in the etalon, and $\delta$ defines the dispersion in the etalon. The $\tau$ can be derived from the geometry of the etalon as: $\tau = \frac{2h}{c}\left(n-\frac{1}{n}\right)$, where $c$ is the speed of light, $n$ is the refractive index of the etalon, and $h$ is its thickness.

Analysis of the interferograms was carried out using a Matlab code with appropriate velocity per fringe (VPF) calibration values to change phase (image) to velocity. The interferometric data was taken with and without the shock wave application for each measurement and former data was used as the background phase.

\subsection{Modelling of shock conditions}

Modelling of shock propagation in black Kapton (polyimide) was numerically modelled using the MULTI software package~\cite{mult,mult2}. 
HEA films were $\sim 1~\mu$m-thick; see Figs.~\ref{f-AuSlice} and \ref{f-sec}. 

In experiments, the strongest modification of the XRD peaks occurred $\sim 4$~ns after the laser pulse, which was irradiated into the 25~$\mu$m-thick black Kapton. The speed of shock wave is estimated $v_{sh}\approx 25~\mathrm{\mu m}/4~\mathrm{ns} = 6.25$~km/s; see experimental Kapton shock wave data~\cite{rus} in Fig.~\ref{f-kapt}, 
The pressure estimated (see Sec.~\ref{Hugoniot}) from the impedance matching Hugoniot analysis was at $\sim 55$~GPa. It constitutes a $\sim 25\%$ fraction from the Young modulus predicted from the first principles for similar materials as investigated in this study, e.g., $E = 217$~GPa for CoCrFeCuNi~\cite{ZHANG2022104059}. 

\section{Results}

\subsection{XRD}

The results from flat panel detector 
are analysed 
in Figs.~\ref{AuHEA} and \ref{FeHEA} for CuPdAgPtAu HEA(Au-HEA) and CrFeCoNiCuMo HEA(Fe-HEA), respectively. Only the face-centered cubic (FCC) phase is observed in both cases. For Au-HEA, lattice planes (111), (200), and (220) show a peak splitting when the time delay between the pump and the probe is $\geq$ 4.0~ns. Unlike Au-HEA, Fe-HEA did not show peak splitting, and the peaks for (111) and (200) have full width at half maximum (FWHM) $>$0.2~\r{A}$^{-1}$ and overlap with each other. This could indicate dislocation of the lattice structure at smaller angles~\cite{ungar_microstructural_2004} or a smaller crystallite structure is formed for Fe-HEA than Au-HEA~\cite{ungar_meaning_2003}. The peak broadening is observed when $\Delta t >$ 4.2~ns in both cases and shifts toward the larger angles (larger $Q$) with increasing $\Delta t$. The maximum compression of Au-HEA estimated from the (111) plane from 3.9\AA~ to 3.7\AA~ corresponds to $5.1\%$ change.

\subsubsection{XRD of AgAuCuPdPt under high pressure}

High-entropy alloys (HEAs) are known for their exceptional structural stability under applied stress, owing to sluggish atomic diffusion and nanoscale lattice distortions. These properties result in high strength and ductility, making HEAs well-suited for demanding applications~\cite{Tsai03}. 
Despite this stability, HEAs may still undergo various nanostructural changes under uniaxial pressure, including phase transitions, lattice parameter shifts, amorphization, and dislocation generation. The structural evolution of the AgAuCuPdPt HEA under shock compression is presented in Fig.~\ref{AuHEA}(b). Prior to compression, the CuPdAgPtAu (Au-HEA) exhibits a face-centered cubic (fcc) crystal structure with a lattice parameter of $a = 3.94$\AA, as confirmed by peak fitting using the Pawley method. The nature of shock-induced structural evolution depends on the magnitude of the applied stress. If the material is compressed within the elastic regime, the changes are expected to be reversible. However, stresses exceeding the elastic limit can lead to irreversible deformation and permanent micro-structural modifications.

The high-pressure compressed AgAuCuPdPt HEA exhibits two trends: 
\blue{compression of initial phase and formation of a transient phase}. During the initial phase of compression, up to 4.0~ns, the lattice parameter remains unchanged, and the XRD pattern shows a single set of Bragg peaks corresponding to the original cubic fcc structure with a lattice parameter of $a = 3.94$\AA. At 4.0 ns, a second set of Bragg peaks emerges in the XRD data, indicating the formation of a new cubic fcc phase with a smaller lattice parameter of $a = 3.69$\AA. From this point onward, two distinct cubic fcc phases coexist: the original phase with $a = 3.94$\AA~ and the newly formed phase with $a = 3.69$\AA. The evolution of these lattice parameters under shock compression is shown in Fig.~\ref{AuHEA}.

Following the peak splitting observed at 4.0~ns, the lattice parameter of the original phase starts to decrease, reaching $a = 3.86$\AA~ at 4.6~ns. In contrast, the newly formed phase, which appears at 4.0~ns with a smaller lattice parameter of $a = 3.69$\AA, shows a trend of increasing period. Its lattice parameter expands over time and eventually converges with that of the original phase $a = 3.86$\AA~ at 4.6 ns. This convergence suggests a re-merging of the two phases or structural homogenization into a new, more compressed state (see discussion in Sec.~\ref{disco}).
 

\begin{figure}[tb]
\centering\includegraphics[width=0.95\textwidth]{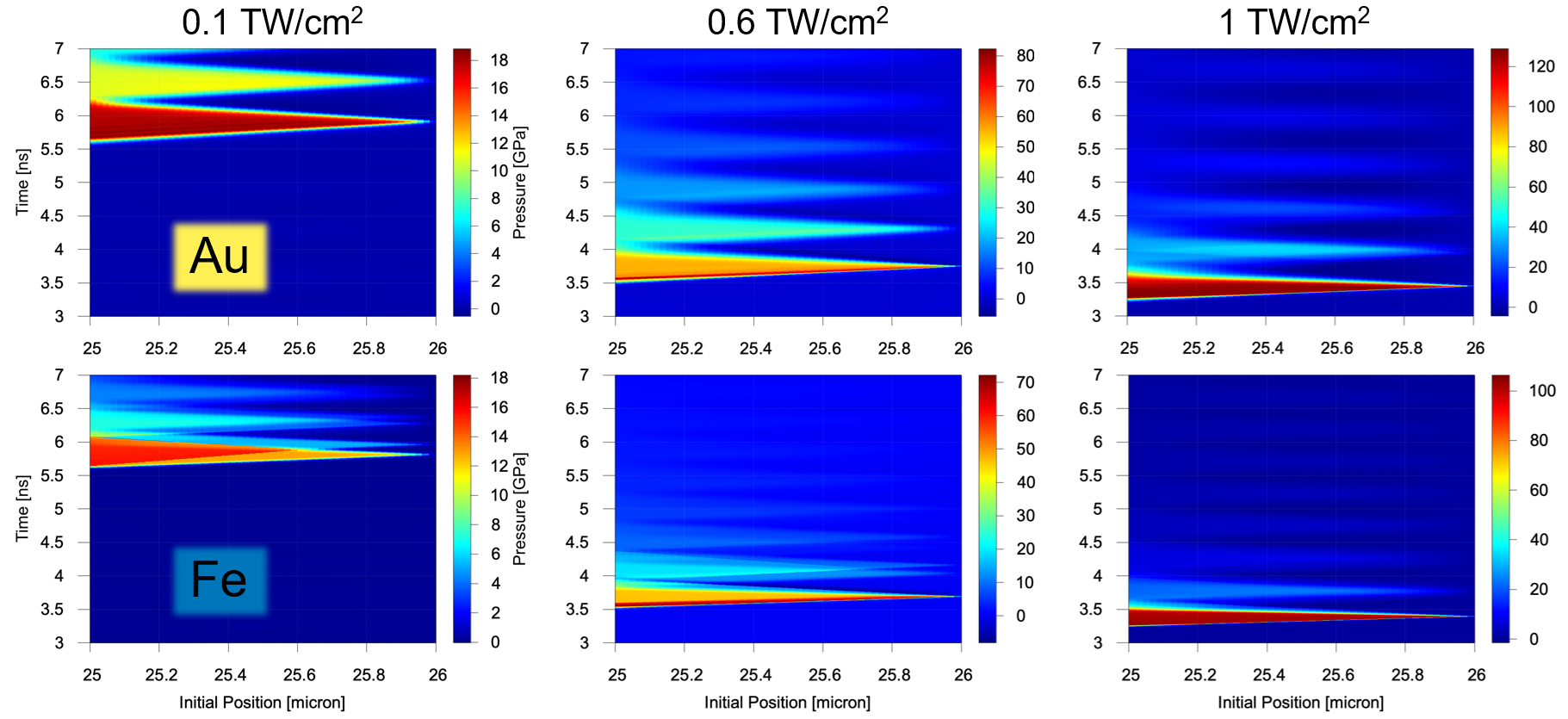}
\caption{Pressure temporal evolution maps for the last 1-$\mu$m thick metal coating (EoS of Au and Fe were used) at different intensities of the laser diving pulse of $0.1, 0.6, 1.0$~TW/cm$^2$ for the used laser and black Kapton ablator. 
Model based on refs.~\cite{mult,mult2} }
\label{f-bar}
\end{figure}
\begin{figure}[tb!]
\centering\includegraphics[width=0.75\textwidth]{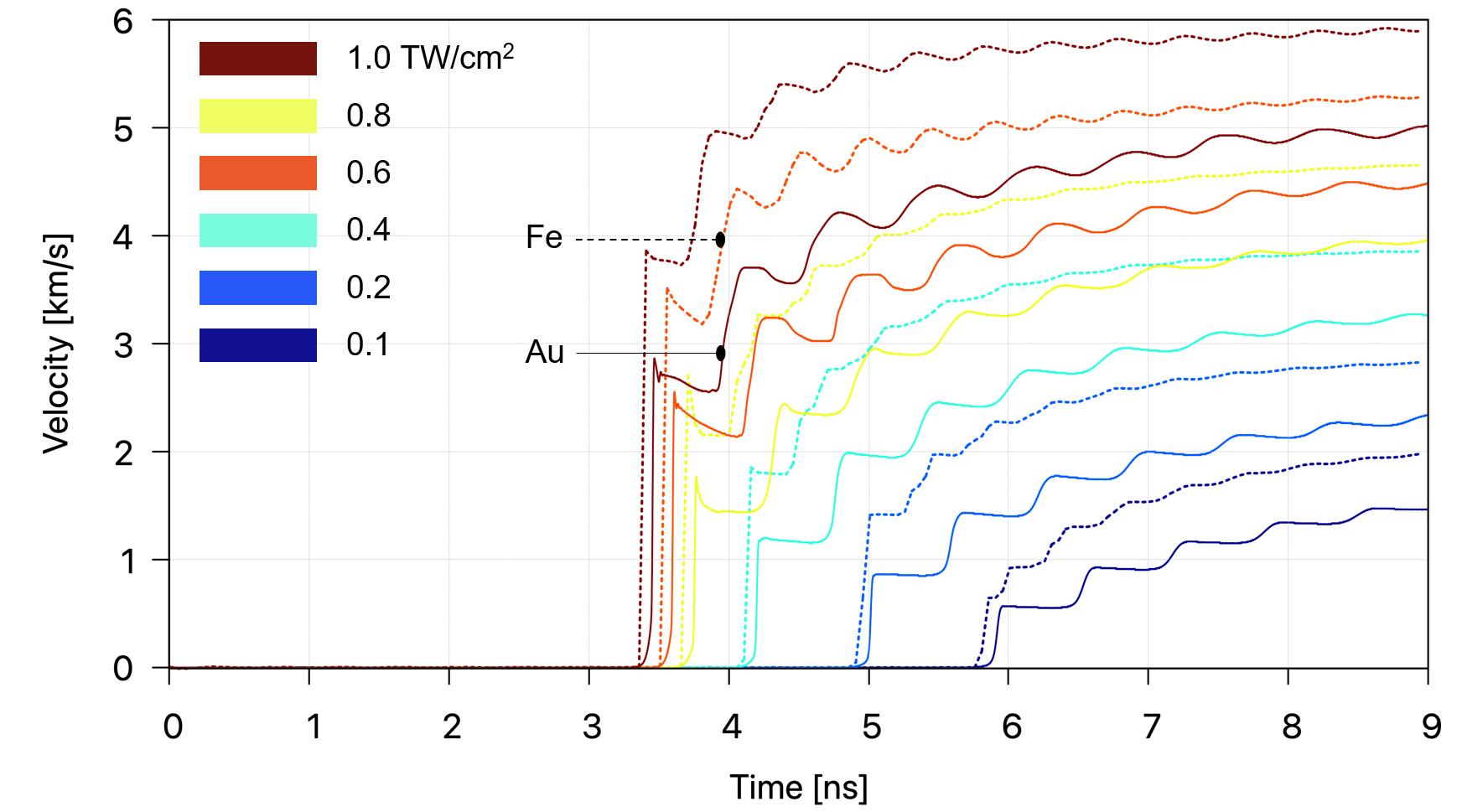}
\caption{The free surface velocity of the rear side of the target Au and Fe at different laser pulse intensities (see reverberations of pressure transients in Fig.~\ref{f-bar} all contributing to the final surface velocity). 
Model based on refs.~\cite{mult,mult2} 
}
\label{f-vis}
\end{figure}

\subsection{VISAR}

VISAR data analysis (Fig.~\ref{f-visar}) shows that shock velocity was larger in Fe-HEA samples as compared with the Fe-HEA coated with AlO$_x$. Also, shock velocity in Au-HEA was larger as compared with the annealed Au-HEA. The shock speed in the black Kapton substrate was in between those for Fe-HEA (larger) and Au-HEA (smaller). For the VISAR2 analysis, a $2\pi$ phase shift had to be added at the starting point of shock wave launch where non-physical negative velocities are obtained following an established procedure~\cite{shift}. The skipped fringes in the interferogram occur where the phase changes more rapidly than the response time of the streak camera. This blind spot/time results in a loss of fringe visibility and continuity while the actual phase may change more than by the one fringe cycle. With this correction of VISAR2 data, comparable velocities between the two cameras were obtained and shown in Fig.~\ref{f-2vis}; the VISAR fringe analysis was carried out at a single pixel resolution. The final free-surface velocity of the front surface of HEA reached 4.5-5~km/s. 

The shock velocity $D$ can be estimated using the Hugoniot relation $D = U/(1-\rho_0/\rho)$, which defines a $\sim 50\%$ densification of Kapton at such conditions (see Fig.~\ref{f-kapt} for polyimide data). The shock pressure estimate in Kapton is $P_{sh} = \rho_0 DU \approx 23\pm 2.5$~GPa ($\rho_0= 1.41$~g/cm$^3$, $U = 2.6$~km/s, $D = 6.25$~km/s, $\Delta U/U = \Delta D/D \simeq 5\%$). The shock pressure traversed into Au and projected onto its free surface as measured by VISAR is discussed next using the shock impedance matching (IM) method~\cite{Damien,acoustic}. 
  
\section{Discussion}\label{disco}

Formation of transient high-density phase 
is discussed next. Working hypothesis is: i) formation of a meta-stable phase of HEA (Fig.~\ref{AuHEA}(c); Sec.~\ref{meta}) occurring together with ii) elastic compression at early interaction times.
The density increase by $\frac{\rho}{\rho_0}\approxeq 1.05$ due to 
\blue{high pressure} compression was experimentally observed by XRD. 
The Young modulus of HEAs is about 220~GPa and the cubic phase with the lattice constant of $3.7$\AA~ was observed. If it is due to an elastic deformation of the original HEA lattice by $\epsilon = (3.9-3.7)/3.9 \approx 0.05$, 
the required stress follows from the Hooke's law $ \epsilon = \frac {\sigma}{E}$: $\sigma=\epsilon E=0.05\times 220$~GPa $\approx 11$~GPa. However, the compressed phase showed expansion, while the initial HEA underwent compression in 3.5-4.5~nm time window. The length of 7~fs (2.1~$\mu$m in free space) X-ray pulse is comparable with the HEA film's (tilted) thickness (Fig.~\ref{AuHEA}(a)). It ``reads'' (by diffraction) the actual state of HEA from 30~$\mu$m radius and 1~$\mu$m long (thickness of HEA film) cylindrical volume. At the very early times, that volume has region not affected by 
\blue{pressure wave}, which later is compressed. Formation of meta-stable phase is postulated due to possible interplay between entropy and enthalpy terms in the Gibbs energy (Sec.~\ref{meta}). This phase is expanding (due to heating) after initial spontaneous formation (Fig.~\ref{AuHEA}(d)). While the initial HEA undergoes compaction, the meta-stable component is expanding and their lattice periods converging up to the 4.6~ns time, which was the limit in these preliminary beamline-trial experiments.  

Analysis of pressure evolution at the implemented conditions is presented next considering 
\blue{pressure transition} through the interfaces~\cite{Damien1}. This is standard approach for the bulk samples ($> 0.1$~mm thickness). 

\subsection{Hugoniot analysis}\label{Hugoniot}

\begin{figure}[tb]
\centering\includegraphics[width=0.95\textwidth]{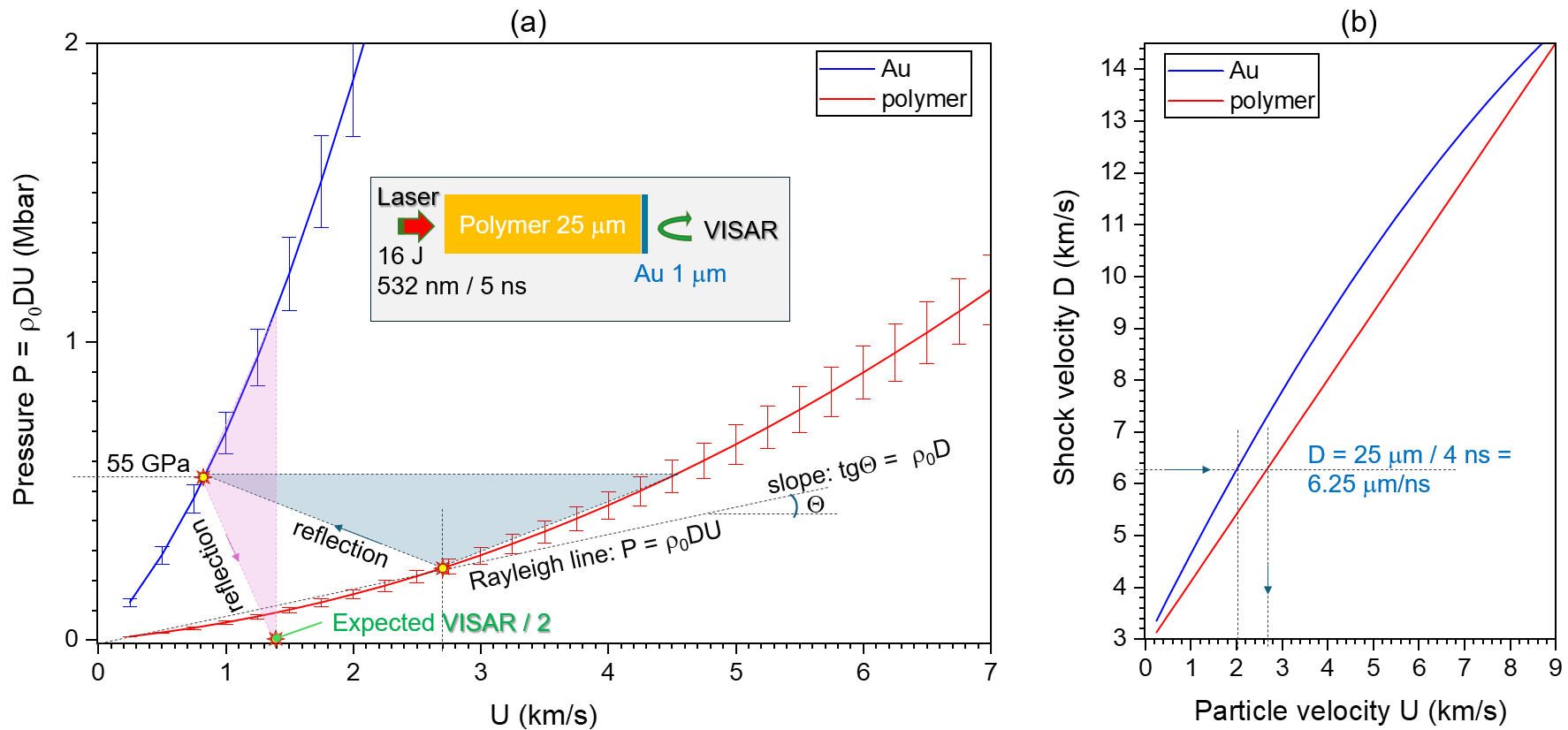}
\caption{The Hugoniot impedance matching (IM) analysis: (a) pressure $P$ vs. particle velocity $U$ and (b) shock\blue{/pressure wave} velocity $D$ vs. $U$. The EoS for Au~\cite{Zhiguo} and polymer (polystyrene and polypropylene)~\cite{Damien} were used. The slope of the Rayleigh line is the shock impedance $\rho_0\times D$. 
Noteworthy, Au was only 1-$\mu$m-thick in this study (inset). This could cause the velocity of Au free surface convoluted with polymer movement measured by VISAR (with $1.4 < D\leq 6.25$~km/s between reflection from Au free surface and shocked polymer). Error bars 10\%.
}
\label{f-shock}
\end{figure}

The standard pressure-velocity $P = \rho_0 UD$ analysis for polymer-metal target is applied (see Fig.~\ref{f-shock}(a)), where EoS of Au~\cite{Zhiguo} and polymer (polystyrene and polypropylene)~\cite{Damien} were used. For Au~\cite{Zhiguo}: $D = 2.910 + 1.775U - 0.051U^2$, with $\rho = 15$~g/cm$^3$ and for polymer~\cite{Damien}: $D = 21 + 1.3(U-14)$ and $\rho = 1.41$~g/cm$^3$; here $D,U$~[km/s]. Figure~\ref{f-shock} shows graphical procedure of acoustic impedance matching (IM) in $P\propto U$ plot for EoS which are qualitatively approximate to those of Au-HEA and black Kapton used in experiments. The experimentally observed shock transit time through a 25~$\mu$m thick Kapton was $\sim 4$~ns. This defines shock velocity $D = 6.25$~km/s in Kapton and particle velocity $U = 2.7$~km using EoS above (Fig.~\ref{f-shock}(b)). For this $U$ value, the IM method was applied to determine the 
pressure $P$ upon 
entrance into Au, which reads 55~GPa. The reflection unloading from free-surface of Au is expected to produce $U \approx 1.4$~km/s, which is half of the surface velocity measured by VISAR behind the 
\blue{pressure} front $V_{VISAR} \approxeq 2U$~\cite{acoustic,Damien}, i.e. the expected VISAR measured velocity is $V_{VISAR} \approx 2.8$~km/s. However, the VISAR velocity of Au-HEA was 4-4.6~km/s (Figs.~\ref{f-visar},\ref{f-2vis}), which is considerably higher than the expected value from IM analysis. A possible explanation is due to a very thin 1~$\mu$m thickness of the film, which only partly contributes to the measured VISAR signal at earliest stages of the movement onset where $V_{VISAR} < 2.5$~km/s (Fig.~\ref{f-visar}). At later stage (8-10~ns; Figs.~\ref{f-visar}) the VISAR signal is dominated with Kapton contribution when approaching maximum saturated VISAR velocities. Indeed, for Kapton $2U_{kap}\approx 5.2$~km (Fig.~\ref{f-shock}(b)) is closer to the experimental $V_{VISAR} = 4.6$~km/s 
as compared to the gold's impedance matching prediction $2U_{Au} = 2.8$~km/s (Fig.~\ref{f-shock}(a)).

The discussion above has a qualitative nature since EoS of Au and polymer were used, which are expected to be slightly different from those of Au-HEA and Kapton. One can expect that the 
pressure in Au-HEA was close to $55\pm 6$~GPa. It is noteworthy, the presented analysis was only qualitative and 
\blue{the pressure} velocity and VISAR data were not matching. This can be caused by thin $\sim 1~\mu$m HEA film, while the method is known to work quantitatively for the \blue{$\geq 20~\mu$m} thick polished samples~\cite{Lin_2025,Parsons_2026}. Moreover, the min-max roughness of the HEA film was $15-20\%$ of the thickness due to non-flat surface of black-Kapton (see schematics in Fig.~\ref{AuHEA}(a)). The roughness is expected to contribute to a higher uncertainty in 
\blue{pressure} transition/reflection from interfaces, especially for thin micro-films~\cite{Karzova}.  

\subsection{Transient phase of HEA}\label{meta}

We propose that the observed and unexpected peak splitting (Fig.~\ref{AuHEA}) indicates that, under 
\blue{a high pressure}compression, the initial Au-HEA phase temporarily evolves into two distinct HEA phases that coexist for a brief period of approximately 0.6~ns. These include the original phase with a larger lattice parameter and a newly formed intermediate phase, which appears to be highly metastable and disappears shortly after the 
\blue{pressure} is removed. The formation of this metastable phase can be rationalized in terms of thermodynamic phase stability under non-equilibrium conditions.

The single-phase AgAuCuPdPt high-entropy alloy solid solution forms as a result of a negative Gibbs free energy of mixing $\Delta$G$_{sol}$, which is enabled by the high configurational entropy $\Delta$S$_{sol}$ associated with its multi-component nature. This relationship is governed by the thermodynamic equation:
\begin{equation}
\Delta G_{sol} = \Delta H _{sol} -T \Delta S _{sol},
\end{equation}
\noindent where $\Delta$G$_{sol}$ is the Gibbs free energy change of the solution (in this case, the AgAuCuPdPt HEA), $\Delta$H$_{sol}$ is the enthalpy of mixing, T is the absolute temperature, and $\Delta$S$_{sol}$ is the configurational entropy. The increase in entropy, $\Delta$S$_{sol}$, helps suppress the formation of competing intermetallic compounds (e.g., A+B, B+C, A+C, etc.) and instead promotes the stabilization of a single-phase random solid solution, ABCDE-solution~\cite{rev,Taiwan}.

The tendency for solid solution formation in multi-component alloys can be quantified using the parameter $\Omega$, defined as:
\begin{equation}
\Omega = \frac{(T\Delta S_{sol})}{|\Delta H_{sol}|}.
\end{equation}
\noindent When $\Omega$ > 1, the formation of a random solid solution is thermodynamically favored over the formation of intermetallic phases. The stability of a solid solution, such as the AgAuCuPdPt, can be disrupted by changes in the enthalpy of mixing, $\Delta$H$_{sol}$. The general expression for the change in enthalpy is:
\begin{equation}~\label{e-why}
\Delta H_{sol} = \Delta U_{sol} + \Delta (pV),
\end{equation}
\noindent where $\Delta$U$_{sol}$ is the change in internal energy, $p$ is the pressure, $V$ is the volume, and $\Delta (pV)$ represents the pressure-volume work. Although heat input, $\Delta Q$, can also affect the system’s energy, in shock compression scenarios, the dominant contributions typically arise from changes in internal energy and volume due to the sudden application of the high pressure.

Under ambient conditions, the AgAuCuPdPt satisfies the criterion $\Omega > 1$, indicating that the entropy term $T\Delta S_{sol}$ dominates over $|\Delta H_{sol}|$, thus favoring a stable single-phase solid solution. However, under dynamic conditions such as 
\blue{high pressure} compression, this balance may shift. The rapid increase in pressure and resulting compression of the lattice can significantly raise $\Delta H_{sol}$, potentially lowering $\Omega$ below 1. As a result, the thermodynamic driving force for single-phase stability may be lost, and the material can become metastable or unstable, allowing for the nucleation of new intermediate phases.

X-ray diffraction analysis reveals that the lattice parameter of the AgAuCuPdPt decreases under 
compression, indicating a rapid structural response to the applied stress. At 4.0~ns after the onset of the 
\blue{high pressure}, the diffraction pattern shows the emergence of a second set of peaks, signifying the coexistence of two distinct phases. This observation suggests that the 
\blue{pressure}-induced compression not only reduces the unit cell volume of the original phase but also drives the formation of a new, transient or metastable phase. The resulting peak splitting provides direct evidence of this structural transformation, likely triggered by the destabilization of the single-phase solid solution as the enthalpy of mixing $\Delta$H$_{sol}$ increases under extreme loading conditions.

The time evolution of this transient phase requires experimental observation by XRD over a longer time span as well as theoretical predictions of composition to be experimentally verified in future experiments. 


\subsection{HEA-ablation pressure, velocity at the critical surface of Kapton}

Let us estimate laser ablation conditions for the pulse energy $E_p= 16$~J, focal spot size area $S_{foc}=1.735\times 10^{-3}$~cm$^2$ for the laser wavelength of $\lambda = 532$~nm at the average intensity $I = 1.84\times 10^{12}$~W/cm$^2$; the cyclic frequency $\omega =3.54\times 10^{15}$~s$^{-1}$ defines the critical density $n_c = 3.94\times 10^{21}$~cm$^{-3}$. Kapton (containing atoms C, O, N, H) has an average ion mass, $M_{av} \approx 2\times 10^{-23}$~g (the mass number $A = 12$, taken arbitrarily 
). 
The critical mass density (assuming single ionisation per atom) then is $\rho_c = M_{av}\times n_c= 8\times 10^{-2}$~g/cm$^3$. 

Absorption in ns regime occurs in the vicinity of the critical density. Assuming high absorption, fast energy equilibration of electrons and ions and quasi-stationarity the laser energy flow (intensity) transforms into energy flow of heated ions: $I \approx \rho_c v_i^3$, here $v_i$ is the velocity of ions. Taking the above numbers, one gets $v_i=6\times 10^6$~cm/s. The pressure at the critical surface driving heat wave gradually transformed into the shock wave then estimates as
 $P\approx \rho_c v_i^2 = 3\times 10^{12}$~erg/cm$^3\equiv 300$~GPa $\equiv$ 3~Mbar.

The pressure at the shock wave 
front propagating through the Kapton gradually decreases due to the dissipation, energy losses on ablation and on expansion. These losses are easily taken even from 1D hydrodynamic code. Hence pressure at the shock wave front entering HEA is much smaller. Ways to improve the estimates: account for absorption coefficient $<1$, ionisation rate (might be more than one), average ion’s mass.
The 16~J per pulse is a huge energy, which is a few times larger than necessary for ablation of the whole Kapton in the focal volume. 


The pressure at the shock wave front propagating through the Kapton gradually decreases due to the heat conduction perpendicular to the laser beam direction (small), ionisation, bond breaking, energy losses on expansion. The volume of Kapton $S_{foc}\times 25~\mu$m contains $N_{kap}= 4.9\times 10^{-6}$~cm$^3\times 10^{23}\sim 4.9\times 10^{17}$~atoms. Sum of losses per atom is $\epsilon_{loss}=\epsilon_{coh}+\epsilon_{ion}+\epsilon_{exp}$, here the cohesion energy $\epsilon_{coh}\approx 4$~eV/atom, ionisation $\epsilon_{ion}\approx 12$~eV/atom, expansion $\epsilon_{exp}\approx 10$~eV/atom, i.e., $\epsilon_{loss}= 26$~eV/atom. Energy spent for the ablation of Kapton volume affected by laser equals to $\sim 2$~J. Therefore, at the used energy of 16~J, the Kapton should be burned through.

Figure~\ref{f-bar} shows modelling results for the shock pressure evolution in Au and Fe (using their known EoS) at three laser intensities at the ablator surface~\cite{mult,mult2}. Figure~\ref{f-vis} summarises the free-surface velocity from the pressure model. Such velocity transients are numerical predictions for the experimental VISAR results. At 1~TW/cm$^2$ intensity, the maximum shock pressure of 100-120~GPa ($\sim 1-1.2$~Mbar) is theoretically predicted in Fe and Au. The IM analysis of shock pressure in Au predicts an approximately twice lower pressure at $U\approx 1.4$~km/s when experimental conditions with Au-HEA were applied (Fig.~\ref{f-shock}(a)). 


\section{Conclusions and Outlook}

Formation of a transient phase 
of FCC lattice in Au-HEA was detected at $\sim 55$~GPa within short 4-4.2~ns time window after laser-induced shock in black-Kapton ablator of 25~$\mu$m thickness using XFEL SACLA BL3 beamline. 
The meta-stable (transient) ``X''-phase 
is hypothesised based on an increase of $\Delta H_{sol}$ (due to pressure; Eqn.~\ref{e-why}), resulting in $\Delta G_{sol}$ to increase as well. It is probable that for some composition ``X'' the condition $\Delta G_{sol} < \Delta G_X$ is no longer satisfied for HEA$_{ABCDE}$ at high pressures. After the pressure is removed, the condition $\Delta G_{sol} < \Delta G_X$ is restored. The intermediate phase is stable as long as pressure is applied. 

The shock-compression conditions were analysed using Hugonoit impedance matching, numerical \emph{ab initio} modelling with EoS of related materials, and analytical ablation pressure estimates for the theoretically achievable highest pressures.

This hypothesis of transient ``X'' phase will be further explored with a higher temporal resolution and over longer time span. In the current introductory (6 hours) beamline experiment, longer delay times were not tested and are planned for the full proposal. Further improvements of thin-film HEA samples via surface polishing of Kapton and/or \emph{in situ} imidisation of mirror-flat kapton precursor resist is planned to achieve a few-nm min-max roughness. This is expected to develop methodology when the impedance matching Hugonoit analysis will become quantitative even for few-micrometer thick HEA films. Such technique would become highly valuable for characterisation of fast expanding family of HEAs subjected to high pressures.

\small\section*{Acknowledgements}

 We acknowledge the technical support of the BL3 team during the 2024A8503 SACLA beamtime proposal on 21 April 2024. SJ is grateful for support via the Australian Research Council Discovery DP240103231 grant. JM acknowledges support via JST CREST (Grant No. JPMJCR19I3) and KAKENHI (No.22H02137). MR was supported via KAKENHI (Grant No. 22K14200) and SAKIGAKE (Grant No. JPMJPR250E). 
 This work was the result of using research equipment shared in MEXT Project for promoting public utilization of advanced research infrastructure (Program for Advanced Research Equipment Platforms, Grant No. JPMXS0450300324). We are grateful to Takeshi Matsuoka for discussion of shock wave experiments and the Australia-Germany collaboration grant UA-DAAD 57652710.






\bibliography{Zotero,xfeltime}

\setcounter{figure}{0}
\makeatletter 
\renewcommand{\thefigure}{A\arabic{figure}}
\section{Appendix}

Direct momentum transfer from light to black Kapton is negligible since the radiation pressure is $I/c = 0.33$~GPa for $I = 10$~TW/cm$^2$ pulse irradiance, while the maximum intensity was below 2~TW/cm$^2$; $c$ is the speed of light.

\begin{figure}[h]
\centering\includegraphics[width=1\textwidth]{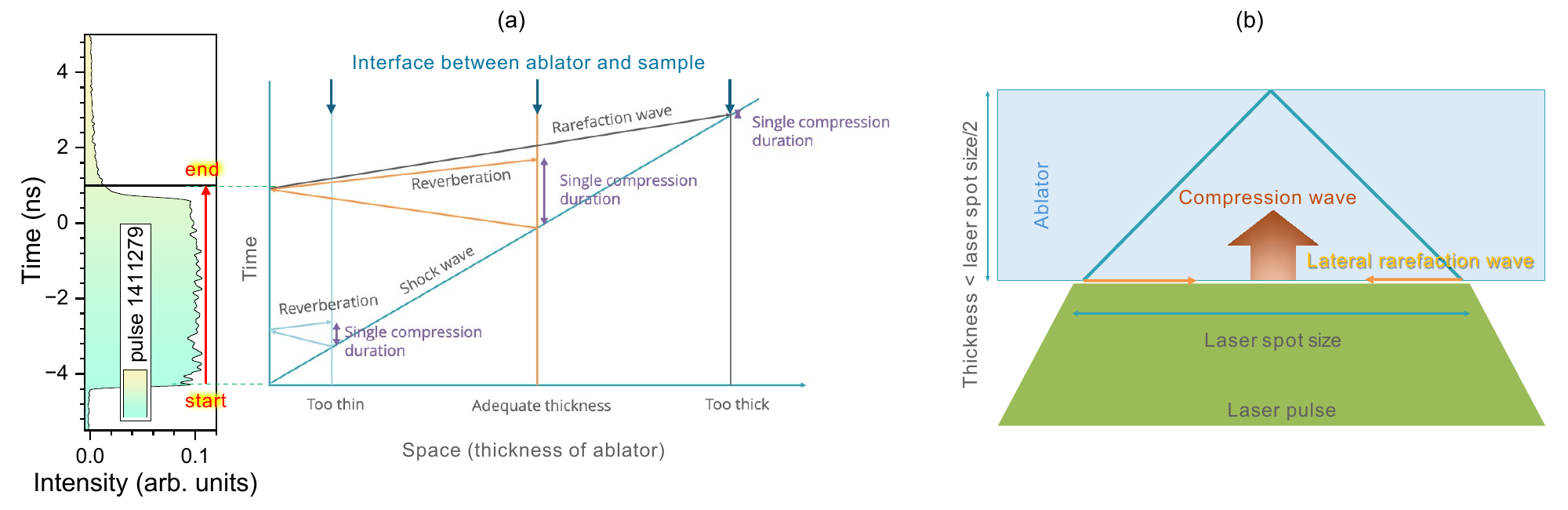}
\caption{Geometrical constraints~\cite{AGRAWAL} on optimal ablator thickness for fixed pulse duration and shock velocity (a) and its relation to the spot size (b). The actual temporal profile of a $\sim$5~ns pulse No.1411279 is shown in (a). Based on a guideline material for experiments at the BL3 beamline. For 6~km/s shock velocity, the rarefaction wave will propagate from the each side of the beam approximately $v_{sh}\times t_{sh}\approx 6~\mu\mathrm{m/ns}\times 4~\mathrm{ns}= 24~\mu$m. } \label{f-cond}
\end{figure}

It is possible to estimate the 
pressure arriving onto the HEA film from the VISAR1 data (Fig.~\ref{f-visar}). A formidable acceleration of $a_{sh} = 7.5\times 10^{12}$~m/s$^2$ is estimated at the onset of surface movement. The mass density of Fe-HEA is $\rho \approx 8.9$~g/cm$^3$ for the equal sum of five constituent metals. 
The underestimated value, we consider the 
\blue{pressure} wave is impacting the same $2r = 470~\mu$m diameter Fe-HEA disk of 1~$\mu$m thickness as on the surface of black Kapton ablator, i.e., the mass of the Fe-HEA disk $m_{HEA} = \rho\times \pi r^2 = 1.54$~ng. This yields in the force $F = m_{HEA}a_{sh} = 11.58$~kN, which exerts the pressure of $F/(\pi r^2) = 66.75$~GPa. This can only serve as an estimate since the shock is concentrated on a slightly reduced $\sim 422~\mu$m diameter 
(Fig.~\ref{f-cond}(b)). 
This estimate is qualitatively consistent with discussion in Sec.~\ref{disco}. We assume that there was no shock induced melting of HEA, which takes place when pressure reach close to the Young modulus of material, e.g., discussed in a similar experiment for gold~\cite{Briggs2019}. 

\begin{figure}[h]
\centering\includegraphics[width=0.7\textwidth]{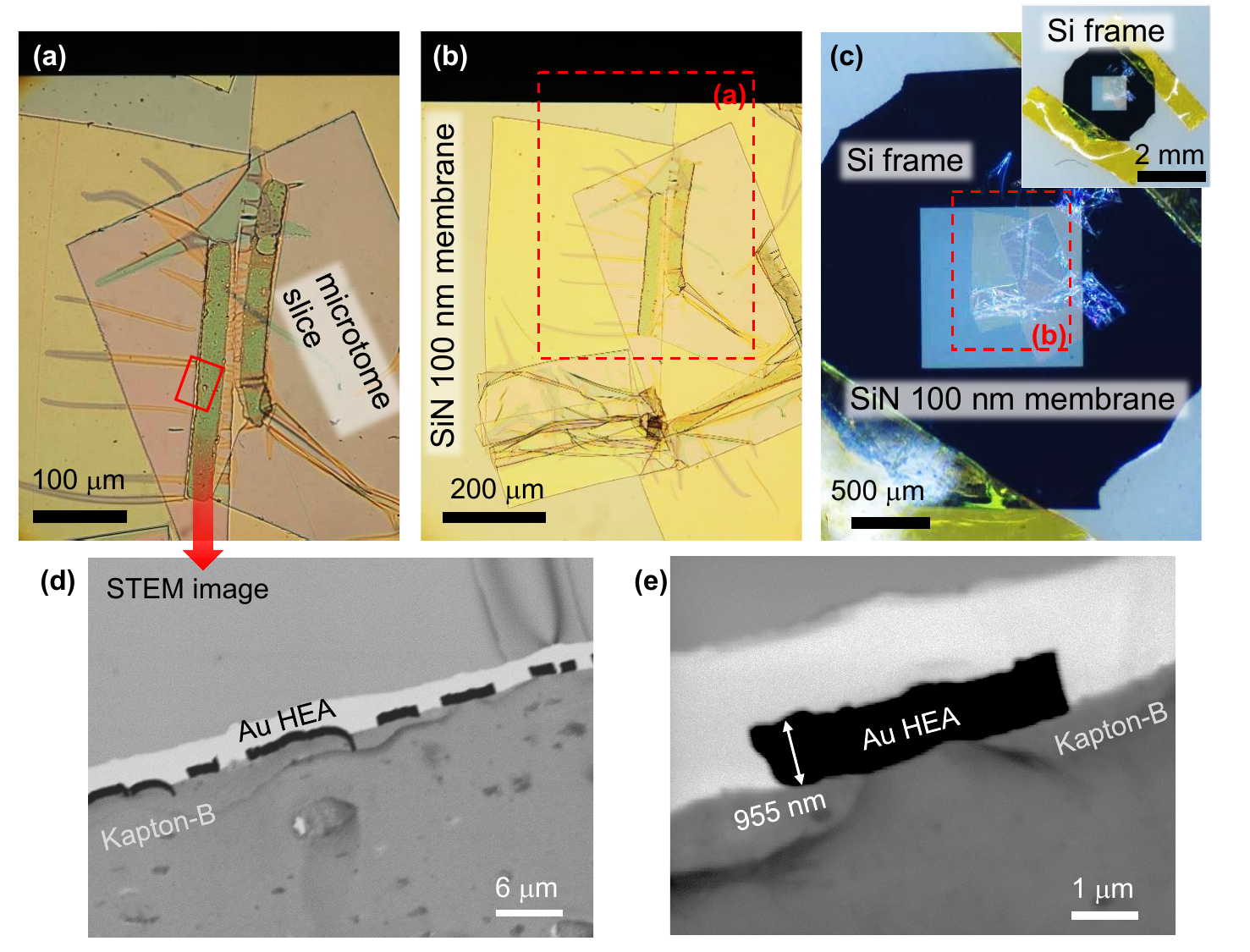}
\caption{Ultra-thin microtome of Au-HEA on black Kapton. (a-c) Optical images of the microtome slice on 100-nm-thick SiN membrane at different magnifications. (d,e) Cross sectional STEM view of the Au-HEA ($960\pm 15$~nm ) on a black Kapton ablator. Thickness of the microtome slice is 90~nm. The sample is from the same batch of samples used in the XFEL experiment.} \label{f-AuSlice}
\end{figure}
\begin{figure}[h]
\centering\includegraphics[width=0.95\textwidth]{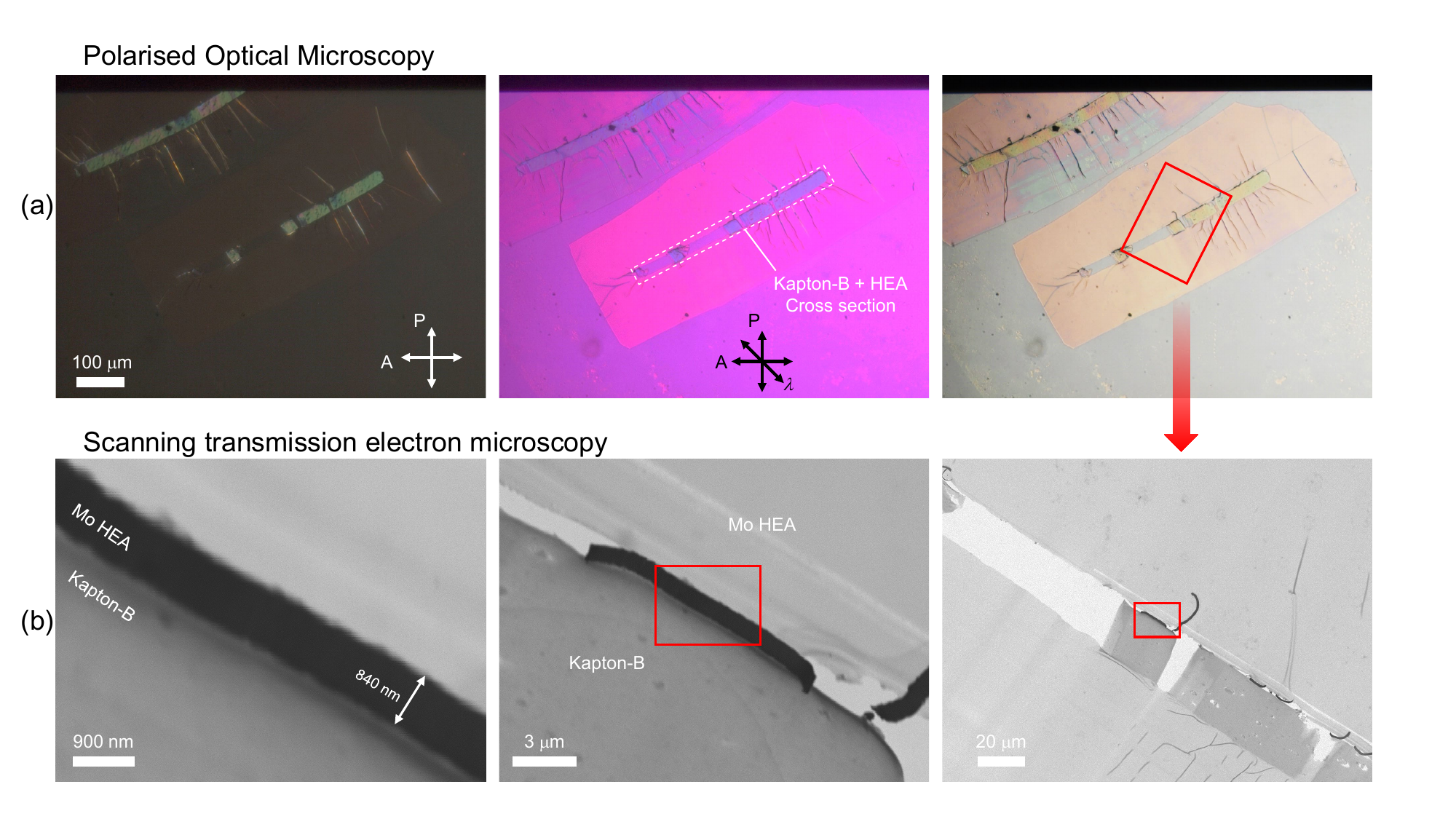}
\caption{Cross sectional view of the Fe-HEA ($840\pm 10$~nm ) on a black Kapton ablator: optical cross polarised imaging (a) and STEM (b). Thickness of the microtome slice is 90~nm. The sample is from the same batch of samples used in XFEL experiment.} \label{f-sec}
\end{figure}
\begin{figure}[h]
\centering\includegraphics[width=1\textwidth]{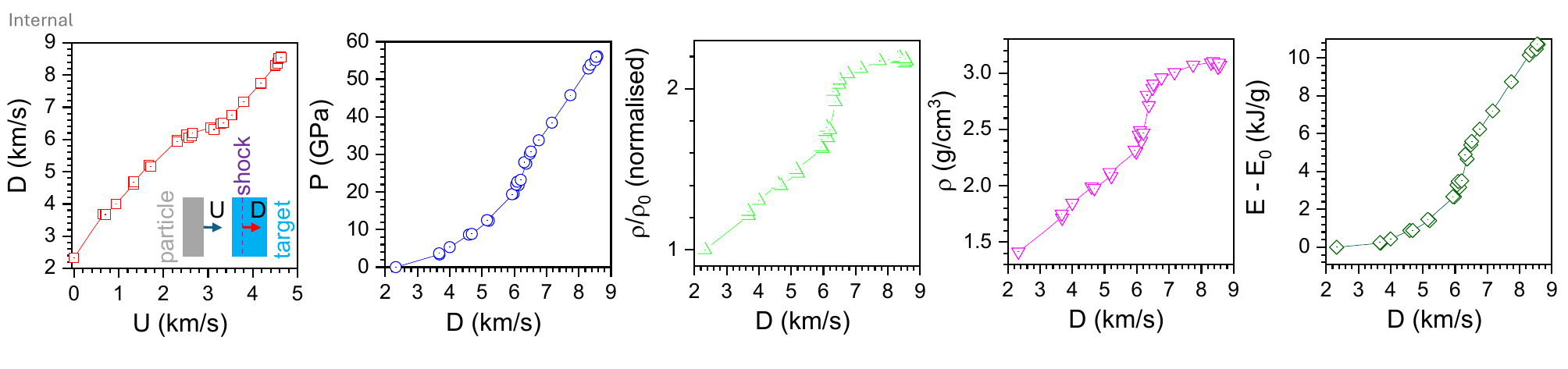}
\caption{Experimental Rankine–Hugoniot shock wave results on polyimide~\cite{rus} (inset), mass density $\rho_0 = 1.414$~g/cm$^3$. Projectile velocity $U$, shock wave velocity in the target $D$, compression ratio behind the shock $\rho/\rho_0$, pressure behind the shock front $P$, density behind shock front $\rho$, and $(E-E_0)=\frac{1}{2}(P_0+P)\left(\frac{1}{\rho_0}-\frac{1}{\rho}\right)$ is the specific energy behind the shock front.} \label{f-kapt}
\end{figure}
\begin{figure}[h]
\centering\includegraphics[width=0.85\textwidth]{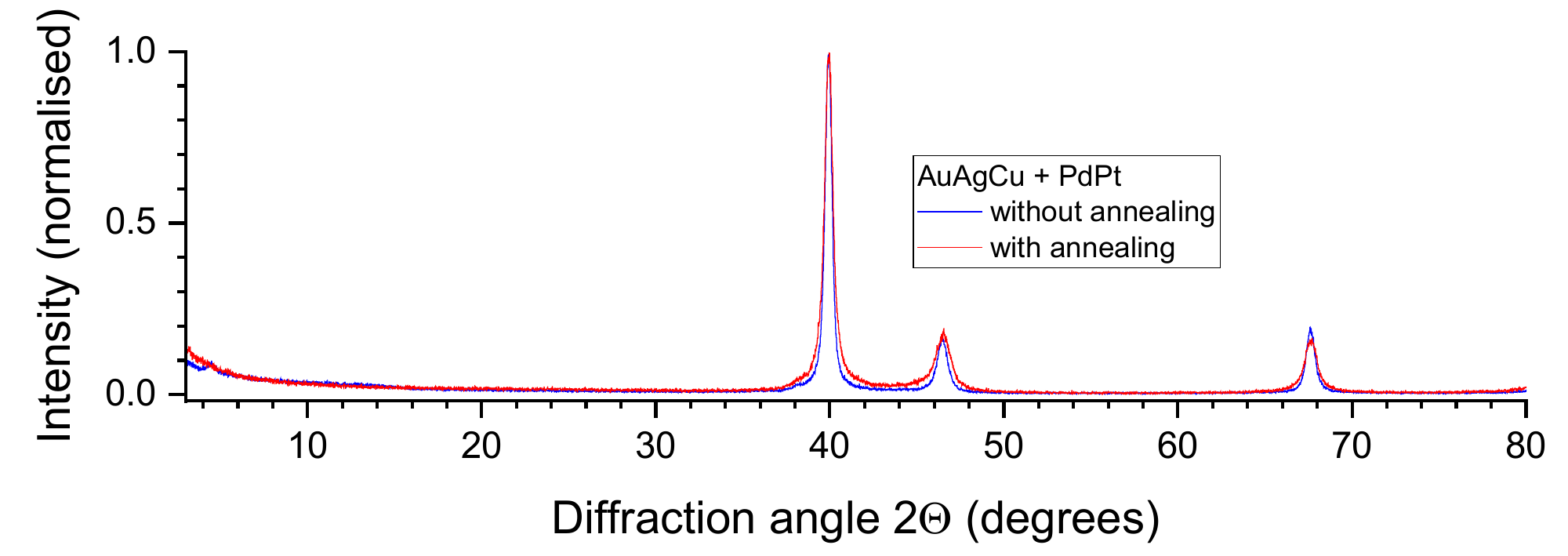}
\caption{Au-HEA $1~\mu$m films were co-sputtered from two targets CuAgAu and PdPt simultaneously. Then one sample was annealed at 150$^\circ$C for 1~hour on black Kapton ablator. Almost identical XRD (Cu-$K_\alpha$) patterns were observed; the main peak is FCC $\left<111\right>$. 
} \label{f-temp}
\end{figure}
\begin{figure}[h]
\centering\includegraphics[width=\textwidth]{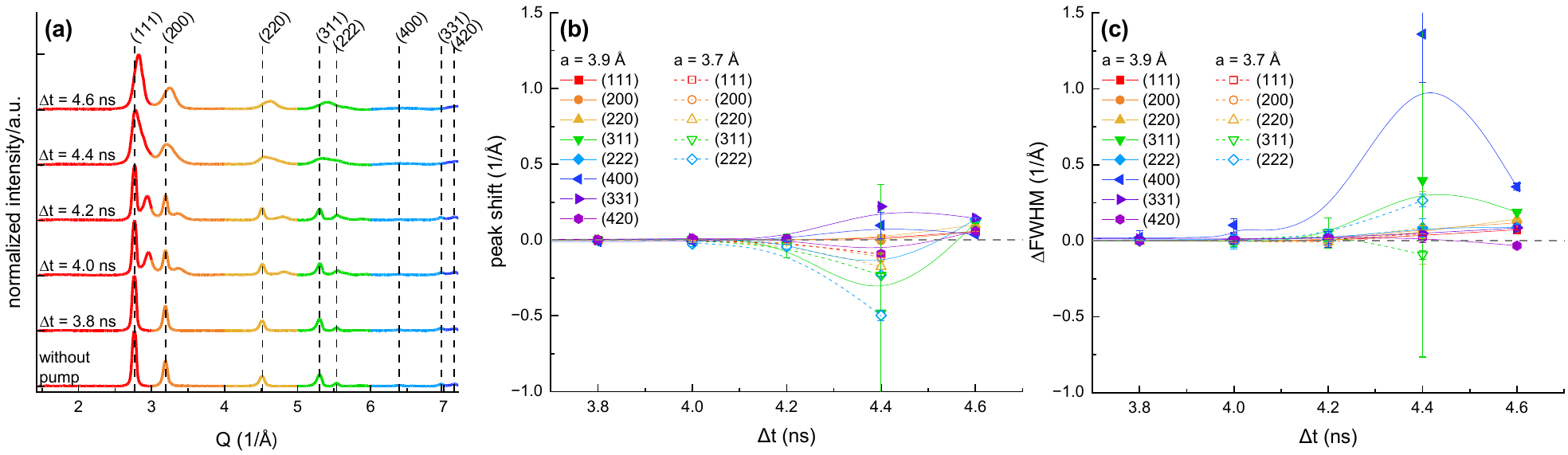}
\caption{The structural evolution upon the application of 
\blue{pressure} wave by the high-speed XRD structural analysis using XFEL for CuPdAgPtAu HEA. The positive peak shift corresponds to compression and negative to dilation.}
\label{f-AuPeaks}
\end{figure}
\begin{figure}[h]
\centering\includegraphics[width=\textwidth]{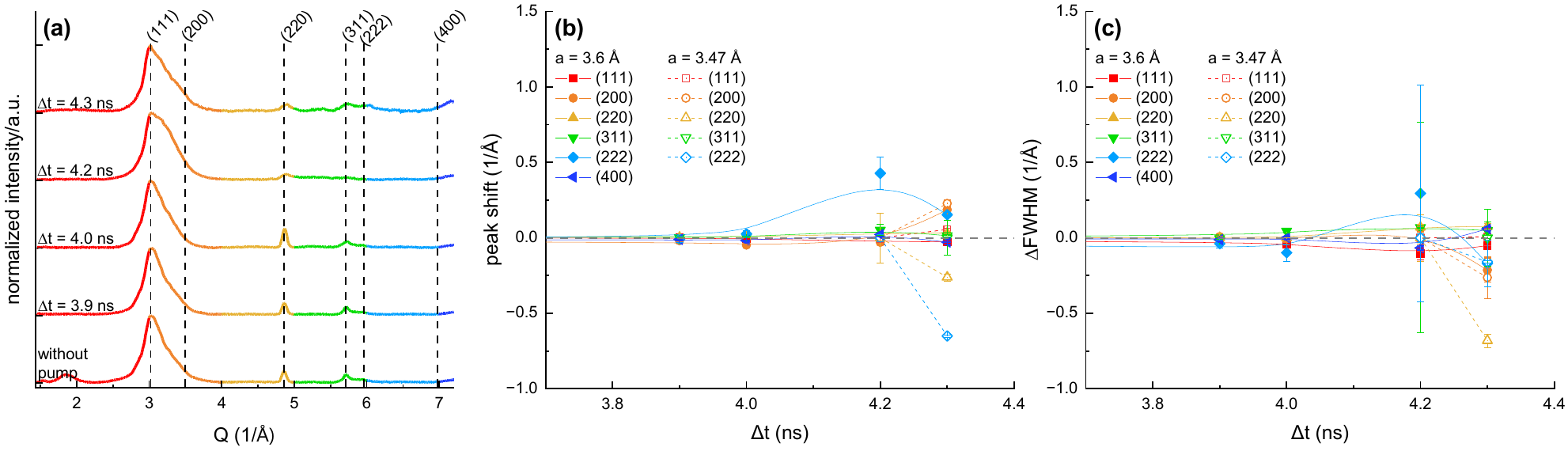}
\caption{The structural evolution upon the application of 
\blue{pressure} wave by the high-speed XRD structural analysis using XFEL for CrFeCoNiCuMo HEA. The positive peak shift corresponds to compression and negative to dilation.} 
\label{f-FePeaks}
\end{figure}
\begin{figure}[h]
\centering\includegraphics[width=0.75\textwidth]{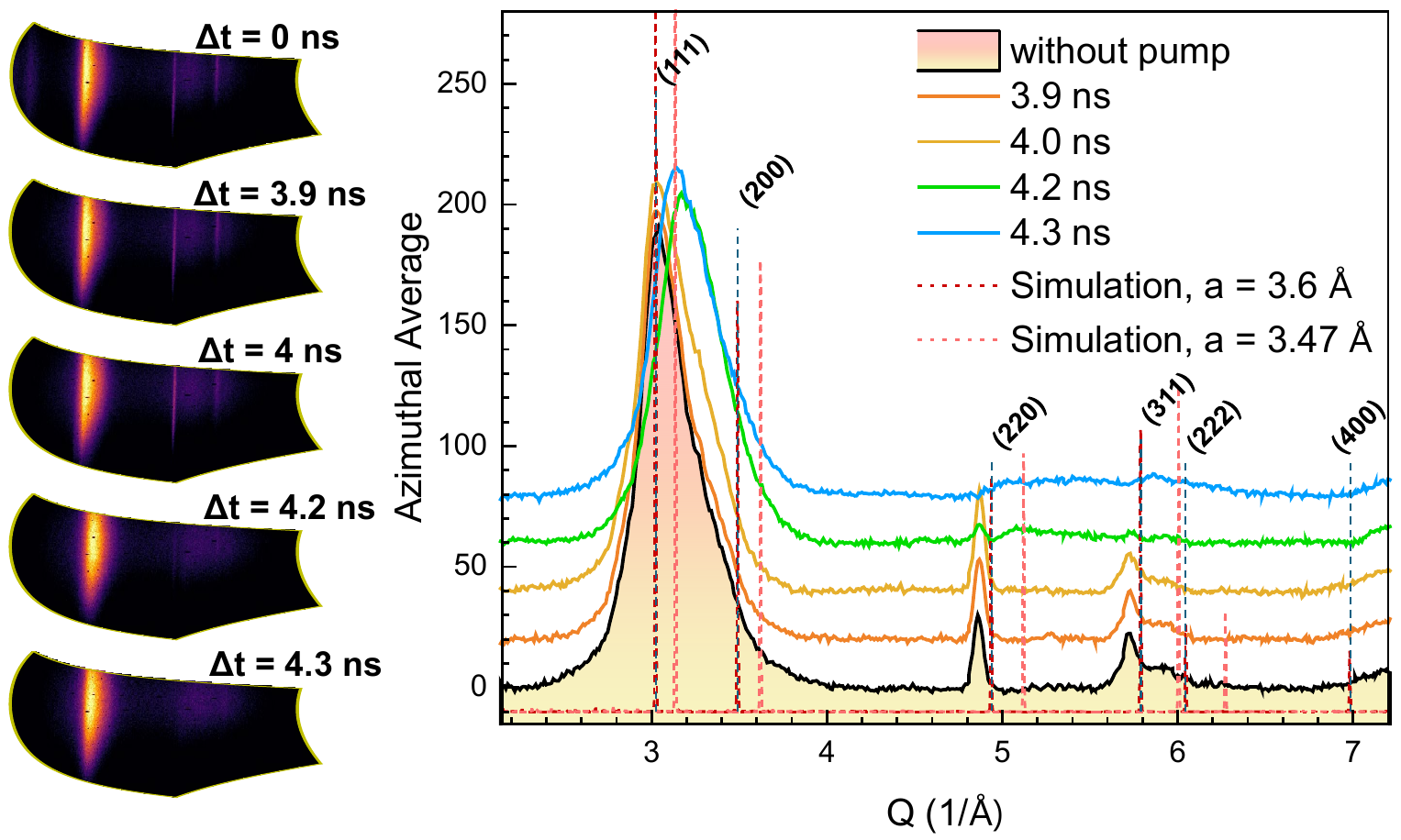}
\centering\includegraphics[width=\textwidth]{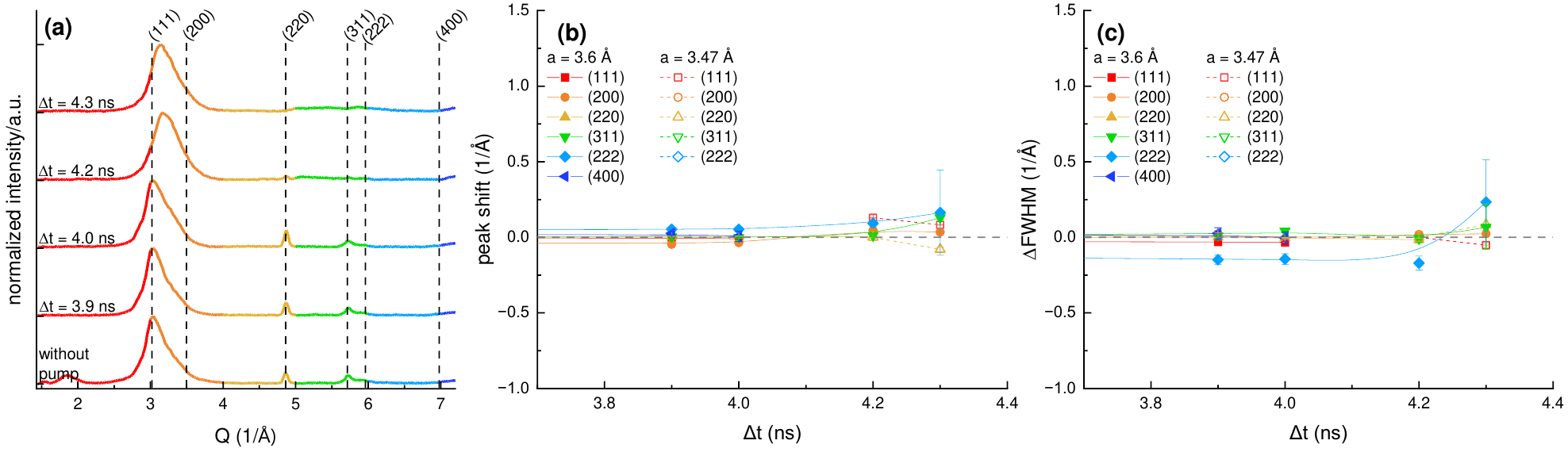}
\caption{The structural evolution upon the application of 
\blue{pressure} wave by the high-speed XRD structural analysis using XFEL for CrFeCoNiCuMo HEA. The positive peak shift corresponds to compression and negative to dilation.}
\label{f-fealxrd}
\end{figure}

\subsection{Geometry and choice of ablator}

Selection of ablator was based on time-space constraints shown schematically in Fig~\ref{f-cond}. A black Kapton ablator of 25~$\mu$m thickness allowed to apply compression for long enough time, which was additionally limited by a relatively small $1~\mu$m thickness of HEA films. In thin films, a back reflected shock from HEA-air interface has $\pi$ phase reversal and reduces initial compression wave. We chose the largest diameter laser beam $470~\mu$m in order to simplify overlap the volume of shocked film and X-ray which was $\sim 30~\mu$m in diameter on front surface of HEA. 

Another constraint on the ablator thickness is due to finite rise time of laser pulse to the intensity maximum (plateau) $t_{up}$. This is important in ablators, e.g., Si, where elastic shock wave has high velocity in a low-pressure region $u_{up}$ (during $t_{up}$). The shock wave launched at the intensity maximum (after rise time) with velocity $u_{top}$ defines the minimum thickness of ablator: $t_{min} = \frac{u_{up}t_{up}}{u_{top}-u_{up}}\times u_{top}$. For Si ablator, $t_{min}\approx 20~\mu$m at 110~GPa. Due to a high longitudinal sound wave velocity $c_L$ (proportional to the elastic wave) in Si along $\left<100\right>$, $\left<110\right>$, $\left<111\right>$ (slow-to-fast), a single shock wave emerges into Si only above 100~GPa at speeds $> 9.5$~km/s~\cite{Si}. The multi-wave structure precludes an ideal single shock compression in samples when Si is employed as an ablator below $\sim 100$~GPa~\cite{Goto1982}.   

\subsection{Rankine–Hugoniot shock data of polyimide}

Experimental Rankine–Hugoniot shock results for Kapton are shown in Fig.~\ref{f-kapt}. 

The sample of Au-HEA was made by sputtering from two targets. Thermal annealing was used to facilitate formation of FCC HEA. XRD with Cu-$K_\alpha$ line showed almost identical structures with and without thermal annealing (Fig.~\ref{f-temp}). 

\subsection{Other samples measured at the same beamtime}

Detailed transient changes of the XRD peak positions of Au-HEA (Fig.~\ref{f-AuPeaks}), Fe-HEA (Fig.~\ref{f-FePeaks}), and \ce{AlOx}-coated Fe-HEA (Fig.~\ref{f-fealxrd}) together with their FWHM for specific planes are obtained with the same baseline subtraction procedures. 




\end{document}